\documentclass[a4paper,12pt]{article}
\usepackage{amsmath}
\usepackage{amsfonts}
\usepackage{mathtools}
\usepackage{authblk}
\usepackage{graphicx,epsfig}
\usepackage{float}
\usepackage{amssymb}
\usepackage{physics}
\usepackage{mathtools}
\usepackage{comment}
\usepackage{soul}
\usepackage{geometry}
\usepackage{color}
\usepackage{array}
\usepackage{enumitem}
\usepackage[unicode,psdextra]{hyperref}
\usepackage[normalem]{ulem}

\numberwithin{equation}{section}
\title{Light Ring Stability Across Spacetime Dimensions and Theories of Gravity}
\author[1,2]{Kajol Paithankar\footnote{kajol.paithankar@iiap.res.in}\,}
\author[1,2]{Sanved Kolekar\footnote{sanved.kolekar@iiap.res.in}\,}
\affil[1]{Indian Institute of Astrophysics, Koramangala II Block, Bangalore 560034, India}
\affil[2]{Pondicherry University, R.V. Nagar, Kalapet 605014, Puducherry, India}

\begin{document}

\maketitle

\begin{abstract}

The light ring theorem states that four-dimensional stationary, axisymmetric, horizonless ultracompact objects (UCOs) possess light rings (LRs) in pairs, one of which is stable if the Null Energy Condition (NEC) holds in General Relativity (GR). The existence of a stable LR may trigger nonlinear spacetime instabilities, thereby raising questions about the viability of UCOs as long-lived black hole mimickers. We extend the LR existence and stability theorem beyond four-dimensional GR to modified gravity theories, in particular, $f(R)$ gravity, scalar-tensor theories, restricted Horndeski gravity and Einstein-Gauss-Bonnet (EGB) gravity, in ($d \geq 4$) dimensional spacetimes. Our findings show that in four-dimensional modified gravities, the stability of LRs is governed by theory-specific coupling parameters in addition to the NEC. Consequently, a stable LR is not guaranteed to occur in certain parameter regimes. In higher dimensions ($d>4$), we confirm using topological arguments that LRs still emerge in pairs; however, the topological charges are now associated with four distinct types of LRs, including two types of saddle points. The stability analysis further constrains the allowed LR pairings, implying that, unlike in four-dimensional GR, the emergent pair does not necessarily include a stable LR. Our findings directly impact tests of GR using UCOs, demonstrating that LR stability is highly sensitive to both the underlying gravitational theory and spacetime dimensionality.
\end{abstract}

\section{Introduction}

Light rings (LRs) are special null trajectories along Killing orbits corresponding to the symmetries of the background spacetime. In suitably adapted coordinates, these invariant structures can take the form of closed circular orbits of photons, resulting from extreme deflection of light in the strong-gravity regime. Around a spherically symmetric black hole, an infinite number of circular null geodesics exist which collectively form a photon sphere. Whereas in stationary, axisymmetric spacetimes of rotating black holes, the number of LRs may be finite. In the case of a Kerr black hole, two LRs exist in the equatorial plane \cite{BardeenKerrorbits}. It has been shown that, in general stationary axisymmetric spacetimes, LRs are located at the extrema of the curves defining the boundaries of the photon region \cite{Kajol}.

The LRs play a central role in interpreting observations from the Event Horizon Telescope (EHT) \cite{M87,SgrA}, the LIGO-Virgo-KAGRA (LVK) collaboration \cite{LIGOfirst}, and future missions such as the Black Hole Explorer (BHEX) \cite{BHEXmotivation}. In the electromagnetic domain, EHT and BHEX observations may reveal a sequence of narrow, self-similar rings formed by photons that execute multiple revolutions in the vicinity of unstable LRs defining the boundary of a black hole shadow \cite{ngEHT,BHEXphoton}. In Kerr spacetime, the prograde and retrograde equatorial LRs correspond to two critical points on the shadow boundary, while the remaining points are associated with spherical photon orbits \cite{TeoSPOs}. These photon subrings provide a probe of spacetime geometry that is largely independent of the properties of the illuminating matter. In particular, their subsequent demagnification is governed by the instability of the LR orbits, providing a characteristic of the underlying geometry \cite{Universalinterferometricsignatures}. Future time-resolved observations from the EHT and BHEX missions will further enable studies of transient phenomena near LRs, such as higher-order images that arrive with specific time delays \cite{Wong2024,Timedelay}. In the gravitational-wave domain, the post-merger ringdown signal detected by LVK encodes the quasinormal modes (QNMs) of the remnant. These QNMs are associated, in the eikonal regime, with perturbations localized near the LRs and their decay into the surrounding spacetime \cite{RingownprobeLR}. These two observables are connected through the shadow-QNM correspondence in the eikonal limit. In particular, the oscillation frequency and the damping rate of QNMs have been associated with the orbital angular velocity and the Lyapunov exponent at the LR \cite{ShadowQNMcardoso}. The shadow parameters are also related to the acceleration bounds of uniformly accelerated trajectories in a static spherically symmetric black hole spacetime \cite{LUAshadowcorrespondence}. Thus, LRs provide a common geometrical link between electromagnetic and gravitational-wave observables of the strong-gravity regime. 

Such extreme gravity is not exclusive to black holes. It can also be generated by black hole mimickers, which are difficult to distinguish from black holes observationally. Black hole mimickers are regular, horizonless ultracompact objects (UCOs) that possess LRs and consequently produce shadow images resembling those of black holes. During the ringdown phase, the gravitational-wave signals emitted by such objects can closely mimic black hole ringdown, while distinctive features associated with their internal structure, such as echoes, may emerge at later stages \cite{RingdownandEchoes}. As a result, a ringdown signal consistent with that of a black hole may provide evidence for the presence of an LR, but does not, by itself, establish the existence of an event horizon or exclude a horizonless UCO. By potentially avoiding event horizons and curvature singularities, these objects challenge the conventional black hole paradigm and motivate the investigation of alternative compact-object models. Several classes of black hole mimickers capable of attaining sufficient compactness to support LRs have been studied, including gravastars \cite{Gravastar}, boson stars \cite{BosonStar}, Proca stars \cite{Procastar}, and wormholes \cite{Wormholes}. It is therefore important to investigate light rings for their universal characteristics and distinctive features across different classes of compact objects and theories of gravity.

An investigation of UCOs as physically viable alternatives to black holes has led to the formulation of the light ring theorem, which establishes constraints on the existence and stability of light rings in the background of a stationary, axisymmetric spacetime describing horizonless UCOs \cite{LRStability}. Using topological arguments based on the Brouwer degree of continuous vector field maps, the theorem demonstrates that any smooth, horizonless UCO formed from continuous gravitational collapse from an approximately flat spacetime must possess light rings in pairs, with one being a saddle point and the other a local extremum of the effective potential. The local extremum further corresponds to a stable light ring if matter obeys the Null Energy Condition (NEC) within the framework of general relativity in $1+3$ dimensions. The presence of a stable light ring has important physical implications; in particular, it allows radiation to pile up nonlinearly, triggering spacetime instabilities. This implies that smooth UCOs are generically unstable and cannot serve as long-lived observational alternatives to black holes \cite{nonlinearinstability}.

 The light ring theorem was subsequently extended to stationary, axisymmetric, asymptotically flat black hole spacetimes, proving that a nonextremal, topologically spherical event horizon enforces the existence of at least one standard (saddle point) light ring outside the event horizon for each sense of rotation \cite{StationaryBHLRs}. Building on the topological approach, the behavior of the potential functions was further analyzed without imposing any restrictions on the object's formation history, and it was demonstrated that nonextremal black holes possess at least two normal (radially unstable and angularly stable) LRs, while at least one retrograde LR exists near extremal black holes \cite{Universalproperties}. For black holes, these existence results were proved in a theory-agnostic way, relying on boundary conditions at the event horizon, at asymptotic infinity, and near the symmetry axis. If the background spacetime possesses an ergoregion, then at least one LR was shown to lie outside the ergoregion \cite{GhoshSarkar2021}. Following the topological argument, LRs were further studied in asymptotically dS/AdS spacetimes \cite{Wei2020}, in toroidal black holes and multi-black-hole spacetimes \cite{blindtotopology} and in wormhole spacetimes \cite{XavierHerdeiroCrispino2024}. It was found that, in the case of spherically symmetric UCOs, the presence of degenerate LRs can allow configurations with an odd number of light rings \cite{HOD2018}.
 
In the formulation of the light ring theorem for UCOs, although the emergence of LRs in pairs was established using theory-independent topological arguments, the proof that one of the two LRs is stable relies on Einstein's field equations. Therefore, the stability result is restricted to UCO solutions of general relativity. This raises the question of whether similar stability properties of LRs hold in modified theories of gravity. To answer this, we investigate the stability of LRs in $1+3$ dimensions in $f(R)$ gravity, Scalar-tensor theories, and restricted Horndeski gravity in Section \ref{stability in MOG}. For each theory, we employ the corresponding field equations and, where possible, exploit general properties of the spacetime to establish definitive criteria of LR stability.

The existing literature discussing the light ring theorem is predominantly focused on four-dimensional spacetime. In section \ref{LRs in HD}, we extend the existence theorem to ($d>4$)-dimensional, stationary, axisymmetric spacetimes with an appropriate circularity condition, assuming only two Killing symmetries. An earlier investigation of LRs in five-dimensional stationary black holes and naked singularities in \cite{LRsin5D} assumed an additional Killing symmetry, with the light ring defined to be the orbit generated by a linear combination of three Killing vectors, yielding an extended light ring structure with spherical rather than circular topology. We investigate the light ring theorem in higher dimensions under less restrictive symmetry assumptions, with spacetime adapted to only two Killing vectors associated with stationarity and axisymmetry, thus retaining the circular topology of LRs as in four-dimensional spacetime. Accordingly, the LR is defined by orbits generated by a linear combination of these Killing vectors, which correspond to circular null orbits in an appropriate coordinate system. We further explore the stability of the LRs in the modified theories of gravity considered in section \ref{stability in MOG}, extending the analysis to $d > 4$ dimensions and additionally in Einstein-Gauss-Bonnet (EGB) gravity.

Our analysis allows us to examine whether the instability associated with stable LRs in UCOs is a universal feature of gravity theories and spacetime dimensions, or instead depends on the underlying gravitational dynamics and dimensionality. We find that in the modified-gravity theories considered here, the existence of a stable LR is not guaranteed even when the NEC is satisfied. This result suggests that the instability argument invoked to question the viability of UCOs in general relativity does not necessarily extend to modified theories of gravity, leaving open the possibility that UCOs may serve as viable alternatives to black holes in these settings.

The paper is organized as follows: In Section \ref{Review of Cunha's results}, we review the light ring theorem in four-dimensional GR. In Section \ref{stability in MOG}, we investigate the stability of LRs in $f(R)$ theory, scalar-tensor theory, and restricted Horndeski theory, and obtain the expressions showing the dependence of stability on the parameters defining these theories. In Section \ref{LRs in HD}, we extend the light ring existence theorem to \(d\)-dimensional spacetimes and examine LR stability in modified theories of gravity, including EGB gravity. The implications are discussed in Section \ref{Discussion}.

The metric signature adopted is $(-,+,+,+)$ and natural units $G=\hbar=c=1$ are used.

\section{Review of the Light Ring Theorem in 4D GR} \label{Review of Cunha's results}

In this section, we briefly review the background setup and the proof of the two aspects of the light ring theorem presented in \cite{LRStability}, namely, LR's existence and stability.

\subsection{Background metric Setup}

Consider a four-dimensional, stationary, and axisymmetric spacetime metric describing a horizonless UCO expressed in standard circular coordinates $(t, r, \theta, \phi)$:
\begin{equation}
ds^2 = g_{tt} dt^2 + 2 g_{t\phi} dt d\phi + g_{\phi\phi} d\phi^2 + g_{rr} dr^2 + g_{\theta\theta} d\theta^2
\end{equation}
The metric is assumed to be invariant under the simultaneous reflection $t \rightarrow -t$ and $\phi \rightarrow - \phi$. The metric components $g_{\mu\nu}$ are smooth, differentiable functions of $r$ and $\theta$, adapted to the Killing vector fields $\partial_t$  and $\partial_\phi$ associated with stationarity and axisymmetry. The cross-terms $g_{r\theta}$ are set to zero via gauge freedom, with $g_{rr} > 0$, $g_{\theta\theta} > 0$, and $g_{\phi\phi} > 0$ to avoid closed timelike curves. The Hamiltonian governing the null geodesic flow of photons is given by
\begin{equation}
H = \frac{1}{2} g^{\mu\nu} p_\mu p_\nu = \frac{1}{2} \left[ K(r,\theta) + V(r,\theta) \right] = 0
\end{equation}
where $p_\mu$ is the photon's four-momentum. The kinetic term $K(r,\theta)$ and potential term $V(r,\theta)$ are separated as
\begin{align}
K(r,\theta) &= g^{rr} p_r^2 + g^{\theta\theta} p_\theta^2 \\
V(r,\theta) &= g^{tt} E^2 - 2 g^{t\phi} E L + g^{\phi\phi} L^2
\end{align}
Here, $E = -p_t$ and $L = p_\phi$ are the conserved energy and angular momentum associated with $\partial_t$ and $\partial_\phi$, respectively. At an LR orbit, $p_r = p_\theta = \dot{p}_\mu = 0$, where the dot represents differentiation with respect to an affine parameter. Using Hamilton's equations, these conditions translate to:
\begin{equation}
V(r,\theta) = 0 \quad \text{and} \quad \partial_i V(r,\theta) = 0 \quad \text{for } i \in \{r, \theta\}
\end{equation}
Further expressing the effective potential in terms of potential functions $H_\pm(r,\theta)$ as
\begin{equation}
V(r,\theta) = -\frac{L^2 g_{\phi\phi}}{D} (\sigma - H_+)(\sigma - H_-)
\end{equation}
where $\sigma = E/L$ and $D = g_{t\phi}^2 - g_{tt} g_{\phi\phi} > 0$, the condition for a LR reduces to finding a critical point where $\partial_i H_\pm = 0$. The potential functions $H_\pm(r,\theta)$ are given by
\begin{equation}
 H_\pm = \frac{-g_{t\phi} \pm \sqrt{D}}{g_{\phi\phi}}  \label{potential functions}
\end{equation}
where the $\pm$ signs correspond to the corotating and counterrotating LRs, respectively.

\subsection{Pairs of LRs}

The topological part of the light ring theorem relies on the Brouwer degree of a vector field map $v_\pm^i = \partial^i H_\pm$. The Brouwer degree is given by:
\begin{equation}
\text{deg}(v_\pm) = \sum_n w_n =\sum_n \text{sgn}\left[ \det\left(\partial_j \partial^i H_\pm\right) \right]_{x_n}
\end{equation}
where $x_n$ represents the location of a LR and $w_n$ is the topological charge evaluated at $x_n$. For a two-dimensional vector field on the $(r,\theta)$ space, a local extremum (maximum or minimum) of $H_\pm$ contributes a topological charge $w = +1$, whereas a saddle point contributes $w = -1$. The Brouwer degree is an invariant topological quantity under continuous deformations of the metric. Hence, for a UCO formed through continuous gravitational collapse from an initially less compact spacetime that doesn't possess LRs, the total topological charge must vanish, and LRs must necessarily occur in pairs of a saddle point ($w = -1$) and a local extremum ($w = +1$).

\subsection{Stability of LRs}

The stability of the local extremum is determined by considering GR as the theory of gravity. At the location of the LR, the NEC can be associated with the second derivatives of the potential through the Einstein field equations as
\begin{equation}
8\pi \rho = 8\pi T_{\mu\nu} p^\mu p^\nu =R_{\mu\nu} p^\mu p^\nu= \frac{1}{2} \partial_i \partial^i V \ge 0
\end{equation}
where index $i$ represents coordinates $(r,\theta)$. In two dimensional space $(r,\theta)$ with $g_{r\theta}=0$, the NEC at LR implies
\begin{equation}
\partial_i \partial^i V = g^{rr} \partial_r^2 V + g^{\theta\theta} \partial_\theta^2 V \ge 0
\end{equation}
Since $g^{rr} > 0$ and $g^{\theta\theta} > 0$, the NEC prohibits both second derivatives from being simultaneously negative ($\partial_r^2 V < 0$ and $\partial_\theta^2 V < 0$). Thus, the local extremum ($w = +1$) must be a local minimum of potential $V$, corresponding to a stable LR.

The topological proof for the emergence of LRs in pairs is independent of the underlying theory of gravity and hence is applicable for a UCO in any theory of gravity in four dimensions. The stability proof, however, uses Einstein's field equations. Therefore, the extremum being a local minimum of the potential holds only for a UCO solution in GR. In the following sections, we test the stability of the LR corresponding to the local extremum in three modified gravity theories.

\section{Stability of LRs in Modified gravity theories} \label{stability in MOG}

In this section, we investigate the LR stability in $(1+3)$ dimensions within modified theories of gravity, namely $f(R)$ gravity, scalar–tensor theories, and Horndeski gravity \cite{MOGReview1, MOGReview2}. For each theory, we employ the corresponding field equations and, where possible, exploit general properties of the spacetime geometry and the NEC to establish criteria for LR stability.

In particular, we obtain a constraint on the second derivatives of the effective potential by relating them to the geometric part of the field equations of the corresponding theory. Following the procedure and convention in \cite{LRStability}, we use indices $(a,b,c,d)$ for Killing coordinates $(t,\phi)$, indices $(i,j,k,l)$ for non-Killing coordinates $(r,\theta)$ and $(\mu,\nu,\rho,\sigma)$ for all spacetime coordinates. Additionally, we define a tensor $\mathcal{M}^{ij}$, which we identify as the factor (other than any numerical factor) contracted with the Hessian of the potential $\mathcal{H}_{ij}=\partial_i\partial_j V$ in the Null energy condition for each theory. This tensor provides the weights corresponding to each component of the Hessian $\mathcal{H}_{ij}$. For GR, one can simply find that $\mathcal{M}^{ij}=g^{ij}$.

\subsection{\texorpdfstring{$f(R)$}{f(R)} Gravity} \label{$f(R)$ Gravity section}

The $f(R)$ theory of gravity generalizes Einstein's general relativity, with the action given by \cite{fRReviewModPhys}
\begin{eqnarray}
    S &=& \frac{1}{2\kappa} \int d^4x \,\sqrt{-g}\; f(R)
\end{eqnarray}
where $g=\text{det}(g_{\mu\nu})$ and $f(R)$ is a function of the Ricci scalar $R$. The $f(R)$ gravity is therefore a family of theories, each one defined by a different function, $f(R)$. The simplest case, $f(R)=R$, is general relativity. Varying the action with respect to the metric, we get the field equations of the theory given as
\begin{eqnarray}
\frac{df}{dR} R_{\mu\nu}-\frac{1}{2} g_{\mu\nu} f(R)+ \left[\,g_{\mu\nu}\, \Box -\nabla_\mu \nabla_\nu \,\right] \, \frac{df}{dR} &=& \kappa \,T_{\mu\nu}
\end{eqnarray}
where $\Box=g^{\mu\nu}\, \nabla_\mu \nabla_\nu$. To simplify the calculations, we divide the left-hand side into three separate tensor contributions as
\begin{equation}
A_{\mu\nu} = \frac{df}{dR} R_{\mu\nu}, \quad B_{\mu\nu} = -\frac{1}{2} g_{\mu\nu} f(R), \quad C_{\mu\nu} = \left[ g_{\mu\nu} \Box - \nabla_\mu \nabla_\nu \right] \frac{df}{dR}
\end{equation}
Imposing the NEC, we get
\begin{equation}
\kappa \rho = \kappa T_{\mu\nu} p^\mu p^\nu = p^\mu p^\nu \left( A_{\mu\nu}+B_{\mu\nu}+C_{\mu\nu} \right) \geq 0
\end{equation}
where $p^\mu$ is any null vector. Now, considering $p^\mu$ to be a tangent vector of a LR, we evaluate the contraction of $p^\mu p^\nu$ with the three terms in the bracket separately at the location of an LR. The contraction in the first term is the same as that in the GR case, with an additional multiplication factor. Hence we get
\begin{equation}
 A_{\mu\nu} p^\mu p^\nu = \frac{df}{dR} R_{\mu\nu} p^\mu p^\nu = \frac{df}{dR} \left( \frac{1}{2} \partial_i \partial^i V \right)   
\end{equation}
Since $p^\mu$ is a null vector, the second term gives
\begin{equation}
    B_{\mu\nu} p^\mu p^\nu = -\frac{1}{2} f(R) \left(g_{\mu\nu} p^\mu p^\nu\right) = 0
\end{equation}
and the third term simplifies to
\begin{equation}
C_{\mu\nu} p^\mu p^\nu = p^\mu p^\nu g_{\mu\nu}\, \Box \frac{df}{dR} - \,p^\mu p^\nu \nabla_\mu \nabla_\nu \, \frac{df}{dR} = - \,p^\mu p^\nu \nabla_\mu \nabla_\nu \, \frac{df}{dR}
\end{equation}
Expanding the covariant derivatives in the above expression, we get,
\begin{equation}
C_{\mu\nu} p^\mu p^\nu=-p^\mu p^\nu \nabla_\mu \nabla_\nu \left(\frac{df}{dR}\right) = -p^\mu p^\nu \left[ \partial_\mu \partial_\nu \left(\frac{df}{dR}\right) - \Gamma^\sigma_{\mu\nu}\, \partial_\sigma \left(\frac{df}{dR}\right) \right]
\end{equation}
At the location of the light ring, since $p^r=p^\theta=0$, we can reduce the expression using the Killing coordinate indices $(a,b)$ as
\begin{equation}
C_{\mu\nu} p^\mu p^\nu = - p^a p^b \left[ \partial_a \partial_b \left(\frac{df}{dR}\right) - \Gamma^\sigma_{ab}\, \partial_\sigma \left(\frac{df}{dR}\right) \right]  
\end{equation}
In a stationary axisymmetric spacetime, the metric functions depend only on the coordinates $(r,\theta)$. As a result, the Ricci scalar and the function $df/dR$ of the Ricci scalar are also functions of only the $(r,\theta)$ coordinates. This implies that the partial derivative of $df/dR$ with respect to the Killing coordinate indices in the above equation vanishes, and we get,
\begin{equation}
C_{\mu\nu} p^\mu p^\nu = - p^a p^b \left[ -\,  \Gamma^\sigma_{ab}\, \partial_\sigma \left(\frac{df}{dR}\right) \right]  
\end{equation}
Further, using the relation $p^a p^b \Gamma^\sigma_{ab} = \frac{1}{2} \partial^\sigma V$, established in \cite{LRStability}, we can write
\begin{equation}
    C_{\mu\nu} p^\mu p^\nu = \frac{1}{2} \partial^\sigma V \;\partial_\sigma \left(\frac{df}{dR}\right)
\end{equation}
Since, by definition $\partial^\sigma V = 0$ at the LR, we get,
\begin{equation}
C_{\mu\nu} p^\mu p^\nu = 0
\end{equation}
Now, combining the contributions from the three terms, at the location of the LR, we have
\begin{equation}
\kappa \,T_{\mu\nu} p^\mu p^\nu = p^\mu p^\nu \, A_{\mu\nu} = \frac{df}{dR} \left( \frac{1}{2} \partial_i \partial^i V \right)
\end{equation}
Then, at LR, the NEC implies
\begin{equation}
\frac{df}{dR} \left( \frac{1}{2} \partial_i \partial^i V \right) \geq 0 \label{Stability condition in fR}
\end{equation}
Now, we can read off the tensor $\mathcal{M}^{ij}$ as $(df/dR) g^{ij}$. Compared to the GR case, here the sign of the second derivatives is constrained additionally by the sign of $df/dR$ evaluated at LR. When $df/dR>0$, the NEC implies $\partial_i \partial^i V\geq 0$, which is the same condition as GR. The $f(R)$ theories with $df/dR>0$ are ghost-free $f(R)$ gravity models \cite{fRLivingReview}. Therefore, in ghost-free $f(R)$ gravity, if the NEC is satisfied at LR, the extremum of $H_{\pm}$ is a local minimum of the potential, and the corresponding LR is stable. For consistency, one can check that the choice $f(R)=R$ gives the stability condition in the GR case.

In case of $f(R)$ theories that have $df/dR<0$, the NEC implies $\partial_i \partial^i V=g^{rr} \partial_r^2 V+g^{\theta\theta} \partial_\theta^2 V\leq 0$. Since $g^{rr}>0$ and $g^{\theta\theta}>0$, NEC is violated if both $\partial_r^2 V$ and $\partial_\theta^2 V$ are positive. In such $f(R)$ theories, the extremum occurs with both $\partial_r^2 V$ and $\partial_\theta^2 V$ being negative, and the corresponding LR is locally unstable when the NEC is satisfied.

\subsection{Scalar-Tensor Theory} \label{Scalar tensor section}

The scalar-tensor theory of gravity is a modified gravity theory in which a scalar field is introduced to describe gravity, giving an additional scalar degree of freedom. Unlike GR, where gravity is described in terms of the metric tensor alone, in scalar-tensor gravity, it arises from the interplay between the metric tensor and the scalar field. In general, in such theories, Newton's gravitational constant $G$ is no longer constant but a quantity that depends on the scalar field. The action of a general scalar-tensor theory of gravity is given as \cite{scalartensortheories}
\begin{eqnarray}
  S &=& \frac{1}{2\alpha} \int d^4x \,\sqrt{-g}\; \left[\Phi \,R-\frac{\mathbf{\omega}(\Phi)}{\Phi} \partial_\sigma\Phi\, \partial^\sigma\Phi-\mathcal{V}(\Phi)\right]
\end{eqnarray}
where $g$ is the metric determinant, $R$ is the Ricci scalar, $\alpha$ is a coupling constant, $\omega(\Phi)$ is the Brans-Dicke parameter and $\mathcal{V}(\Phi)$ is the scalar-field potential. The variation of the action with respect to the metric gives the field equations as
\begin{eqnarray}
R_{\mu\nu}-\frac{1}{2} R g_{\mu\nu} &=& \frac{\alpha}{\Phi} T_{\mu\nu} +\frac{1}{\Phi} \left[\nabla_\mu \nabla_\nu-g_{\mu\nu} \Box \,\right] \Phi \nonumber\\
& & +\frac{\omega(\Phi)}{\Phi^2} \left[\partial_\mu \Phi \, \partial_\nu \Phi-\frac{1}{2} g_{\mu\nu}\, \partial_\sigma \Phi \,\partial^\sigma \Phi \right]-\frac{g_{\mu\nu}}{2\Phi} \mathcal{V}(\Phi)
\end{eqnarray}
To connect the geometric part of the field equation to the potential at LR, we contract the above tensor equation with null vectors $p^\mu p^\nu$. As $g_{\mu\nu} \,p^\mu p^\nu =0$, the equation reduces to
\begin{eqnarray}
R_{\mu\nu} \,p^\mu p^\nu &=& \frac{\alpha}{\Phi} T_{\mu\nu}\,p^\mu p^\nu +\frac{1}{\Phi} \,p^\mu p^\nu\,\nabla_\mu \nabla_\nu \, \Phi +\frac{\omega(\Phi)}{\Phi^2} \,p^\mu p^\nu\,\partial_\mu \Phi \, \partial_\nu \Phi
\end{eqnarray}
At the location of LR, the only nonzero components of the null vector $p^\mu$ are $p^t$ and $p^\phi$, represented with indices $(a,b)$. Then the above equation at LR becomes
\begin{eqnarray}
R_{ab} \,p^a p^b &=& \frac{\alpha}{\Phi} T_{ab}\,p^a p^b +\frac{1}{\Phi} \,p^a p^b\,\nabla_a \nabla_b \, \Phi +\frac{\omega(\Phi)}{\Phi^2} \,p^a p^b\,\partial_a \Phi \, \partial_b \Phi
\end{eqnarray}
Now, following the argument as in the case of $f(R)$ gravity, the only independent coordinates in the stationary axisymmetric spacetimes are $(r,\theta)$, which implies that the LHS of the above equation is a function only of $(r,\theta)$. This further implies that the RHS and the field $\Phi$ should also be functions only of $(r,\theta)$. Therefore, the partial derivatives of the field $\Phi$ with respect to the Killing coordinates $\partial_a \Phi$ vanish. Further, as shown for the function $df/dR$, the specific combination $p^a p^b\,\nabla_a \nabla_b \, \Phi (r,\theta)$ is proportional to $\partial^\sigma V$ and hence gives zero contribution at the LR. Altogether, at LR, the only nonzero contribution comes from the first term on the RHS, and we get
\begin{eqnarray}
R_{ab} \,p^a p^b &=& \frac{\alpha}{\Phi} T_{ab}\,p^a p^b
\end{eqnarray}
Now, imposing NEC and using the connection between $R_{ab} \,p^a p^b$ and the second derivatives of potential, we can write
\begin{eqnarray}
    T_{ab}\,p^a p^b &=& \frac{\Phi}{\alpha} \, R_{ab} \,p^a p^b= \frac{\Phi}{\alpha} \left( \,\frac{1}{2} \partial_i \partial^i V \,\right) \geq 0 \label{Stability condition in Scalar Tensor}
\end{eqnarray}
Here, the tensor $\mathcal{M}^{ij}$ is $(\Phi/\alpha) g^{ij}$. Therefore, in scalar tensor theory, the stability of the LR corresponding to the extremum of $H_\pm$ depends on the sign of $\Phi/\alpha$ at the LR. If $\Phi/\alpha>0$, the LR is locally stable, and if $\Phi/\alpha<0$, then the LR is locally unstable. Notably, LR stability is entirely independent of the remaining theoretical parameters, specifically the Brans-Dicke parameter $\omega(\Phi)$ and the scalar-field potential $\mathcal{V}(\Phi)$.

\subsection{Horndeski's theory} \label{Horndeski section}

Horndeski's theories are the most general theories of gravity in four dimensions that lead to second-order equations of motion. These theories are constructed by higher-derivative generalizations of the scalar-tensor theories of gravity. The action for Horndeski's theory is given as \cite{Horndeski1974}
\begin{eqnarray}
    S &=& \int d^4x \,\sqrt{-g}\; \frac{1}{\kappa} \sum_{i=2}^5 \mathcal{L}_i (g_{\mu\nu},\Phi)
\end{eqnarray}
with the Lagrangian densities given by
\begin{eqnarray}
    \mathcal{L}_2 &=& G_2 (\Phi, X)\nonumber\\
    \mathcal{L}_3 &=& -G_3 (\Phi, X) \, \Box\Phi\nonumber\\
    \mathcal{L}_4 &=& G_4 (\Phi, X)\, R+ G_{4X} (\Phi, X) \left[(\Box \Phi)^2-(\nabla_\mu \nabla_\nu \Phi)^2\right]\nonumber \\
    \mathcal{L}_5 &=& G_5 (\Phi, X) \,G_{\mu\nu} \,\nabla^\mu \nabla^\nu \Phi\nonumber\\
    & &-\frac{1}{6} G_{5X} (\Phi, X) \left[(\Box \Phi)^3-3\, \Box\Phi\,(\nabla_\mu \nabla_\nu \Phi)^2+2\,(\nabla_\mu \nabla_\nu \Phi)^3\,\right] \quad \nonumber
\end{eqnarray}
where $G_2$ to $G_5$ are generic functions of scalar field $\Phi$ and $X$, with $X=-1/2\, g^{\mu\nu} \nabla_\mu \Phi \nabla_\nu \Phi$, $G_{iX}$ is derivative $dG_i/dX$, $R$ is the Ricci scalar and $G_{\mu\nu}$ is the Einstein tensor. The functional form of the Horndeski Lagrangian has been restricted using the simultaneous observations of gravitational waves GW170817 from a binary neutron star merger and their electromagnetic counterpart, the gamma-ray burst GRB 170817A \cite{GW170817Paper1,GW170817Paper2,GW170817Paper3,HorndeskiGW170817} to,
\begin{eqnarray}
    \mathcal{L} &=& G_2 (\Phi, X)+ G_3 (\Phi, X) \, \Box\Phi +G_4 (\Phi)\, R \label{restricted Horndeski}
\end{eqnarray}
We use this form of the Lagrangian for further evaluations of the light ring stability analysis. Varying the action with respect to the metric then gives the field equations \cite{HorndeskiEOM} as
\begin{eqnarray}
    \sum_{i=2}^4 G^{(i)}_{\mu\nu} &=& \kappa \,T_{\mu\nu}
\end{eqnarray}
where 
\begin{eqnarray}
    G^{(i)}_{\mu\nu} &=& R^{(i)}_{\mu\nu}-\frac{1}{2} R^{(i)} \,g_{\mu\nu}
\end{eqnarray}
with
\begin{eqnarray}
    R^{(2)}_{\mu\nu} &=& -\frac{1}{2} G_{2X} \,\nabla_\mu \Phi \nabla_\nu \Phi-\frac{1}{2} g_{\mu\nu} \,\left( G_{2X} X-G_2\right)\\
    R^{(3)}_{\mu\nu} &=& \frac{1}{2}\,G_{3X} \bigg\{ \Box \Phi \,\nabla_\mu \Phi \nabla_\nu \Phi+2\, \nabla_{(\mu}X \, \nabla_{\nu)} \Phi+g_{\mu\nu} \, X\, \Box \Phi\bigg\} + G_{3\Phi}  \,\nabla_\mu \Phi \nabla_\nu \Phi\\
    R^{(4)}_{\mu\nu} &=& G_4 R_{\mu\nu}-G_{4\Phi} (\nabla_\mu \nabla_\nu \Phi+\frac{1}{2}g_{\mu\nu}\Box\Phi)-G_{4\Phi\Phi} (\nabla_\mu\Phi\nabla_\nu \Phi-X g_{\mu\nu})
\end{eqnarray}
Then the contraction with the null vector gives
\begin{eqnarray}
     T_{\mu\nu} p^\mu p^\nu &=& \frac{1}{\kappa} \sum_{i=2}^4 G^{(i)}_{\mu\nu}\, p^\mu p^\nu \label{Horndeski FE contraction}
\end{eqnarray}
The contraction with null vectors significantly simplifies the RHS of the above equation as $g_{\mu\nu} p^\mu p^\nu =0$, and we get
\begin{eqnarray}
    \kappa\, T_{\mu\nu} p^\mu p^\nu  &=& R^{(2)}_{\mu\nu} p^\mu p^\nu + R^{(3)}_{\mu\nu} p^\mu p^\nu + R^{(4)}_{\mu\nu} p^\mu p^\nu
\end{eqnarray}
Expanding the terms and using $g_{\mu\nu} p^\mu p^\nu =0$, we get 
\begin{eqnarray}
    R^{(2)}_{\mu\nu} p^\mu p^\nu &=& -\frac{1}{2} G_{2X}\, p^\mu p^\nu \,\nabla_\mu \nabla_\nu \Phi\\
    R^{(3)}_{\mu\nu} p^\mu p^\nu &=& \frac{1}{2}\,G_{3X} \left[ \Box \Phi \,\nabla_\mu \Phi \nabla_\nu \Phi+\nabla_{(\mu}X \, \nabla_{\nu)} \Phi\right] \, p^\mu p^\nu + G_{3\Phi} \, p^\mu p^\nu \,\nabla_\mu \Phi \nabla_\nu \Phi\\
    R^{(4)}_{\mu\nu} p^\mu p^\nu &=& G_4\, p^\mu p^\nu\,R_{\mu\nu}-G_{4\Phi} \,p^\mu p^\nu\,\nabla_\mu \nabla_\nu \Phi-G_{4\Phi\Phi}\,p^\mu p^\nu\, \nabla_\mu\Phi\nabla_\nu \Phi
\end{eqnarray}
Now, considering the equation at LR, the nonzero contribution comes from $(\mu,\nu)=(t,\phi)$ represented by $(a,b)$. Further, for simplification, we assume the scalar field functions $\Phi$ and $X$ to be functions only of $(r,\theta)$ as it naturally arises in the case of Scalar-tensor theories (for the dynamical fields refer to \cite{Babichev2014Dressing}). This leads to the vanishing contribution from terms $\nabla_a \Phi$, $\nabla_a X$ to the RHS of Eq.(\ref{Horndeski FE contraction}). Moreover, as derived in the previous sections, for $\Phi=\Phi(r,\theta)$,
\begin{eqnarray}
    p^a p^b \nabla_a \nabla_b \Phi &=&  - p^a p^b \,\Gamma^\mu_{ab} \partial_\mu \Phi=-\frac{1}{2} \partial^\mu V \, \partial_\mu \Phi
\end{eqnarray}
which vanishes at the LR. Then, the only nonzero contribution to the RHS of Eq.(\ref{Horndeski FE contraction}) at LR is given by
\begin{eqnarray}
    T_{\mu\nu} p^\mu p^\nu &=& \frac{1}{\kappa} G_4(\Phi)\, R_{ab} \,p^a p^b
\end{eqnarray}
Then using the relation between $ R_{ab} \,p^a p^b$ and second derivatives of potential, NEC implies
\begin{eqnarray}
    T_{\mu\nu} p^\mu p^\nu &=&  \frac{1}{\kappa} G_4(\Phi)  \left( \,\frac{1}{2} \partial_i \partial^i V \,\right) \geq 0 \label{Stability condition in Horndeski}
\end{eqnarray}
Therefore, in the restricted form of Horndeski's theory given by Eq.(\ref{restricted Horndeski}) wherein the field $\Phi$ depends only on $(r,\theta)$, the tensor $\mathcal{M}^{ij}=G_4(\Phi) g^{ij}$ and the stability of the LR corresponding to the extremum of $H_\pm$ depends on the sign of $G_4(\Phi) $ evaluated at LR. If $G_4(\Phi) >0$, the LR is locally stable, and if $G_4(\Phi) <0$, then the LR is locally unstable.

Note that deriving a concise NEC relation, as in Eq.(\ref{Stability condition in Horndeski}), is specific to the restricted Horndeski theory and does not generalize to the full Horndeski Lagrangian. The field equations of the full theory contain terms of the form $R_{\mu\alpha\nu\beta}\nabla^\alpha\Phi\nabla^\beta\Phi$ which, on contraction with $p^\mu p^\nu$, do not reduce to a structure directly proportional to the second derivatives of the potential at the LR. It remains an open question whether introducing any further assumptions could simplify the NEC relation in the full Horndeski theory.

In this section, we investigated the LR stability in four-dimensional stationary axisymmetric spacetimes in three modified gravity theories. In the next section, we reconstruct the theorem in dimensions $d>4$. In higher dimensions, since the vector field $v_\pm^i = \partial^i H_\pm$ is not two-dimensional, the local extremum and the saddle point cannot be simply associated with the topological charges $+1$ and $-1$, and a proper identification of LR categories is required to establish the theorem.

\section{Light Ring Theorem in Higher Dimensions} \label{LRs in HD}

We now extend the light ring existence and stability theorem of ultracompact objects to generic $d$-dimensional, circular, stationary, axisymmetric spacetimes with \(d>4\). The derivation closely follows that of \cite{LRStability}. For completeness, we present the essential steps of the analysis. Despite the close similarity in the derivation, the resulting conclusions in higher dimensions differ from those obtained in $(1+3)$ dimensions, as discussed at the end of this section.

For generality, we consider only two symmetries of the UCO spacetime that are associated with the two Killing vectors $\partial_t$ and $\partial_\phi$ as in the case of four dimensions. The remaining $(d-2)$-dimensional subspace is defined by the non-Killing coordinates. We assume the spacetime to be circular, wherein the surfaces spanned by the Killing vectors are everywhere orthogonal to the non-Killing directions. We define an LR as a null geodesic orbit generated by a linear combination of the two Killing vectors. We consider the metric to be
\begin{eqnarray}
    ds^2 &=& g_{tt} dt^2 + 2 g_{t\phi} dt d\phi + g_{\phi\phi} d\phi^2 + g_{ij} dx^i dx^j
    \label{metricSA}
\end{eqnarray}
where $x^i\equiv\{x^1,x^2.....,x^{d-2}\}$. The metric functions $g_{\mu\nu}$ are functions of coordinate $x^i$ and are independent of $t$ and $\phi$. For ease of reference and consistency, we label Killing and non-Killing coordinates with separate indices as $x^\mu=\{x^a,x^i\}$, where $\{x^a\}=(t,\phi)$ are Killing coordinates, $\{x^i\}=(x^1,x^2.....,x^{d-2})$ are non-Killing coordinates. The indices $(\mu,\nu,\rho,\sigma)$ are used for all spacetime coordinates, $(a, b, c, d)$ for the Killing coordinates $(t,\phi)$, and $(i, j, k,l)$ for non-Killing coordinates $(x^1,x^2.....,x^{d-2})$. With this notation, we can write $g_{ai}=0$ and $\partial_a g_{\mu\nu}=0$.

The Hamiltonian for null geodesics can be separated into the kinetic and potential parts, given by
\begin{eqnarray}
    H &=& \frac{1}{2} (K+V)= \frac{1}{2} \left(g^{ij} p_i p_j+g^{ab} p_a p_b\right)
\end{eqnarray}
where the kinetic part is defined on a $(d-2)$-dimensional subspace. The momentum components along the Killing directions are conserved quantities, and hence $ p_a = \text{constant} $.

By construction, the tangent to the LR lies in the two-dimensional subspace spanned by the Killing vectors $\partial_t$ and $\partial_\phi$. Consequently, in these adapted coordinates, the LR orbit has circular topology.  It is straightforward to check that at the LR, $p_i=0$ and $\dot{p_i}=0$, where the dot denotes the derivative with respect to an affine parameter. This translates to the following constraints on the effective potential:
\begin{eqnarray}
    V=0 \quad & \text{ and } & \quad \partial_i V=0 
\end{eqnarray}
where the second condition is obtained from Hamilton's equations, and it includes $(d-2)$ independent equations. Following the same procedure as in the four-dimensional spacetime, the potential $V(x^i)$ can be expressed in terms of functions $H_{\pm}(x^i)$ given in Eq.(\ref{potential functions}), and LR can be found to lie at the critical point of $\partial_i H_{\pm}=0$. Hence, in $d$-dimensional spacetime, the location of LR is given by $(d-2)$ non-Killing coordinate values obtained as a solution to the $(d-2)$ equations $\partial_i H_{\pm}=0$.

Further, evaluating the stability of an LR requires analyzing the second derivatives of the effective potential $V$ at LR, which is related to the functions $H_\pm$ through the relation,
\begin{eqnarray}
  \partial_\mu^2 V &=& \pm \left( \frac{2L^2}{\sqrt{D}} \right) \partial_\mu^2 H_\pm
\end{eqnarray}
Since the multiplying factor $2L^2/\sqrt{D}$ is always positive at LR, the sign of $\partial_\mu^2 V$ matches that of $\partial_\mu^2 H_+$ at corotating LR, while it is opposite to the sign of $\partial_\mu^2 H_-$ at counterrotating LR. Hence, when the function \( H_- \) attains a maximum value along a specific direction \( x^\mu \), the associated potential will simultaneously reach its minimum along the same direction. Conversely, if the potential is maximum at LR in a given direction, then the function \( H_- \) will be minimum at LR along that direction. We use these stability properties below while associating the topological charges with LRs.
 
In a two-dimensional $(r,\theta)$ space, as the local maxima and minima of the potential share identical topological charges, the emergence of LRs in pairs can be concluded without differentiating between a maximum and a minimum of potential. In higher-dimensional spacetimes, however, maxima and minima of potential can carry distinct topological charges. This increased complexity makes it essential to distinguish the extrema of the potential and appropriately associate them with their respective potential functions, $H_\pm$.

\subsection{Pairs of LRs in Higher Dimensions} \label{Pairs in HD}

We now choose the vector field map $v_\pm^i = \partial^i H_\pm (x^i)$ on the $(d-2)$-dimensional subspace and define the Brouwer degree of the map
\begin{equation}
\text{deg}(v_\pm) = \sum_n w_n =\sum_n \text{sgn}\left[ \det\left(\partial_j \partial^i H_\pm\right) \right]_{x_n}
\end{equation}
with $x_n$ being the location of an LR. The topological charge at each LR is given by the sign of the determinant of a $(d-2)$-dimensional matrix constructed with elements $\left(\partial_j \partial^i H_\pm\right)$. Using the fact that the determinant is equal to the product of eigenvalues, we now classify each LR into the following four categories based on the specific properties of the eigenvalues of the matrix $\left(\partial_j \partial^i H_\pm\right)$: 
\begin{enumerate}[label=\textbf{\Alph*.},itemsep=1pt]
\item All eigenvalues are positive
\item All eigenvalues are negative
\item An even number of eigenvalues are negative
\item An odd number of eigenvalues are negative
\end{enumerate}
Before assigning a topological charge based on this classification, we first establish how these categories physically relate to LR stability via the potential function. For a corotating LR, if all eigenvalues of $\left(\partial_j \partial^i H_+\right)$ are positive (negative), the LR is locally stable (unstable). Furthermore, the presence of an even (odd) number of negative eigenvalues indicates that the potential reaches a maximum along an even (odd) number of principal directions within the $(d-2)$-dimensional subspace. Conversely, for a counterrotating LR, a spectrum of entirely positive (negative) eigenvalues corresponds to a locally unstable (stable) orbit. In this case, an even (odd) count of negative eigenvalues signifies that the potential minimizes along an even (odd) number of directions in the $(d-2)$-dimensional subspace.

Using the categories, we now assign an appropriate topological charge to each LR based on the sign of the determinant in the $(d-2)$-dimensional subspace and whether $d$ is odd or even. If $d$ is even, the LR of type \textbf{A} or \textbf{B} or \textbf{C} contributes a topological charge $w=+1$, while the LR of type \textbf{D} contributes a topological charge $w=-1$. If $d$ is odd, then the LR of type \textbf{A} or \textbf{C} contributes a topological charge $w=+1$, while the LR of type \textbf{B} or \textbf{D} contributes a topological charge $w=-1$.

In the $1+3$ dimensional spacetime, where the subspace of $\{x^i\}= (r,\theta)$ is two-dimensional, the LRs of types \textbf{B} and \textbf{C} are equivalent, and together with LRs of type \textbf{A}, they all fall under the case of the local extremum; while LRs of type \textbf{D} correspond to the case of a saddle point. However, in spacetime with dimensions $d>4$, LRs of types \textbf{A} and \textbf{B} are local extrema (minimum and maximum, respectively) of $H_\pm$, while LRs of types \textbf{C} and \textbf{D} are saddle points, which we will refer to as saddle type I and saddle type II, respectively.

As the Brouwer degree is conserved under continuous deformation, a UCO formed via the continuous gravitational collapse of an initially LR-free matter distribution must maintain a vanishing total topological charge. Consequently, light rings must emerge in pairs with opposite charges ($w = +1$ and $w = -1$). In even-dimensional spacetimes, a saddle type II ($w = -1$) pairs with either an extremum or a saddle type I ($w = +1$). In odd-dimensional spacetimes, a maximum or saddle type II ($w = -1$) pairs with either a minimum or a saddle type I ($w = +1$). The classifications, maximum, minimum, and extremum, refer strictly to the topology of the functions $H_\pm$ rather than the effective potential itself. We establish the physical stability of these LR pairs with respect to the effective potential in the subsequent section.

\subsection{Stability of LRs in Higher Dimensions} \label{Stability in HD}

In this subsection, we investigate LR stability in the modified theories of gravity considered in Section \ref{stability in MOG}, extending the analysis to spacetimes with $d>4$ dimensions.

As established in Section \ref{stability in MOG}, the stability proof of an extremum LR is dependent on the relation between the contraction $R_{\mu\nu}p^\mu p^\nu$ and the second derivatives of the effective potential $V(x^i)$. Since the remaining steps in the derivation involve tensorial manipulations that are independent of the spacetime dimensionality, they carry over to higher dimensions. We therefore proceed to establish the relation between the second derivatives of the potential $V(x^i)$ and the Ricci tensor contraction $R_{\mu\nu}p^\mu p^\nu$, evaluated at the LR.

Using the definition of potential term $V=g^{ab} p_a p_b$ where $p_a$ is a conserved quantity, we can write
\begin{eqnarray}
    \partial^\mu V &=& p_a p_b \,\partial^\mu g^{ab}= -p^a p^b \partial^\mu g_{ab}= 2 \,p^a p^b \Gamma^\mu_{ab}
\end{eqnarray}
Then the second derivative with respect to coordinates $x^i$ becomes
\begin{eqnarray}
    \frac{1}{2}\partial_i \partial^i V &=& p^a p^b \,\partial_i  \Gamma^i_{ab} +  \Gamma^i_{ab}\, \partial_i \left(p^a p^b\right) \label{second derivative of potential with two terms}
\end{eqnarray}
Now we evaluate the contraction of the Ricci tensor with null vectors $p^\mu p^\nu$ at LR and use the LR constraints $V=0$ and $\partial_i V =0$ to relate it to the above potential derivative terms. At LR, since $p^i=0$, the contraction of the Ricci tensor with the vectors $p^\mu p^\nu$ becomes
\begin{eqnarray}
    R_{\mu\nu} p^\mu p^\nu &=& R_{ab} p^a p^b =  p^a p^b \left(\, \partial_\mu \Gamma^\mu_{ab} -\partial_b \Gamma^\mu_{a\mu}+ \Gamma^\mu_{\nu \mu} \Gamma^\nu_{ab} - \Gamma^\mu_{\nu b} \Gamma^\nu_{a \mu}\,\right)
\end{eqnarray}
Since the metric is independent of the Killing coordinates $\{x^a\}$, the second term in the bracket $\partial_b \Gamma^\mu_{a\mu}=0$ and the first term is nonzero only for $\mu=i$. Further, the third term on contraction with $p^a p^b$ gives
\begin{eqnarray}
  \Gamma^\mu_{\nu \mu} \,\Gamma^\nu_{ab} \, p^a p^b &=& \Gamma^\mu_{\nu \mu} \left(\frac{1}{2}\partial^\nu V\right)
\end{eqnarray}
which vanishes at LR, as $\partial^\nu V=0$. Therefore, at LR, we get
\begin{eqnarray}
    R_{ab}\,p^a p^b &=& p^a p^b\,\partial_i \Gamma^i_{ab} -p^a p^b\,\Gamma^\mu_{\nu b} \Gamma^\nu_{a \mu} \label{Ricci contraction with two terms}
\end{eqnarray}
It was stated in \cite{LRStability} that the second terms in Eqs. (\ref{second derivative of potential with two terms}) and (\ref{Ricci contraction with two terms}) are related by a numerical factor. For completeness, we reproduce its derivation here, including the intermediate steps.

Now, to simplify the second term in Eq. (\ref{Ricci contraction with two terms}), we use the expansion of Christoffel symbols. In the chosen stationary axisymmetric spacetime, the nonzero components of the Christoffel symbols are
\begin{eqnarray}
    \Gamma^i_{ab} &=& -\frac{1}{2} g^{ij} \partial_j g_{ab}\\
    \Gamma^a_{bi} &=& \frac{1}{2} g^{ac} \partial_i g_{bc}\\
    \Gamma^i_{jk} &=& \frac{1}{2} g^{il} \left[\partial_j g_{kl}+\partial_k g_{jl}-\partial_l g_{jk}\right]
\end{eqnarray}
Using these, the second term in Eq.(\ref{Ricci contraction with two terms}) becomes
\begin{eqnarray}
    p^a p^b\,\Gamma^\mu_{\nu b} \Gamma^\nu_{a \mu} &=&  p^a p^b\, \left(\Gamma^i_{c b} \Gamma^c_{a i}+\Gamma^d_{j b} \Gamma^j_{a d}\right)\\
    &=&  \frac{1}{2}p^a p^b\, \left( \Gamma^i_{c b}  \, g^{cd} \partial_i g_{ad} +g^{dc} \partial_j g_{bc} \,\Gamma^j_{a d}  \right)
\end{eqnarray}
As $p_d$ is a conserved quantity, we can write
\begin{equation}
    0= \partial_i \,p_d= \partial_i (g_{ad} p^a)=p^a\,\partial_i (g_{ad} )+g_{ad}\,\partial_i (p^a)
\end{equation}
Using this, we replace the terms of the form $p^a\,\partial_i (g_{ad} )$ with $-g_{ad}\,\partial_i (p^a)$, to get
\begin{eqnarray}
    p^a p^b\,\Gamma^\mu_{\nu b} \Gamma^\nu_{a \mu} &=&  -\frac{1}{2}\, \left(p^b\, \Gamma^i_{c b}  \, g^{cd} g_{ad}\,\partial_i (p^a) +p^a\,g^{dc} g_{bc}\,\partial_i (p^b) \,\Gamma^j_{a d}  \right)
\end{eqnarray}
Since $g^{cd} g_{ad}=\delta^c_a$, the terms in the bracket simplify to
\begin{eqnarray}
    p^a p^b\,\Gamma^\mu_{\nu b} \Gamma^\nu_{a \mu} &=&  -\frac{1}{2}\, \left(p^b\, \Gamma^i_{a b} \,\partial_i (p^a) +p^a\,\partial_j (p^b) \,\Gamma^j_{a b}  \right)\\
    &=& -\frac{1}{2}\,\Gamma^i_{a b}\,\left(p^b \,\partial_i (p^a) +p^a\,\partial_i (p^b) \,\right)\\
    &=& -\frac{1}{2}\,\Gamma^i_{a b}\,\partial_i (p^a\,p^b)
\end{eqnarray}
Then Eq.(\ref{Ricci contraction with two terms}) becomes
\begin{eqnarray}
    R_{ab}\,p^a p^b &=& p^a p^b\,\partial_i \Gamma^i_{ab} +\frac{1}{2}\,\Gamma^i_{a b}\,\partial_i (p^a\,p^b)
\end{eqnarray}
Further, expanding the contraction over the indices $(a,b)$ in terms of individual metric functions, the expression was shown to vanish at LR in four dimensions. We now show that this continues to hold true in any higher-dimensional stationary axisymmetric spacetime with only two Killing symmetries. Using the expression of the Christoffel symbol $\Gamma^i_{ab}$, we get
\begin{eqnarray}
     \Gamma^i_{ab}\, \partial_i \left(p^a p^b\right) &=& -\frac{1}{2} g^{ij} \partial_j g_{ab} \,\left(p^b \,\partial_i (p^a) +p^a\,\partial_i (p^b) \,\right)
\end{eqnarray}
Now using $p^a\,\partial_i (g_{ab} )=-g_{ab}\,\partial_i (p^a)$, we can write
\begin{eqnarray}
    \Gamma^i_{ab}\, \partial_i \left(p^a p^b\right) &=& g_{ab} g^{ij} (\partial_i p^a) (\partial_j p^b)
\end{eqnarray}
 We expand the terms with Killing indices $(a,b)$ in the components along $(t,\phi)$ as
\begin{eqnarray}
    \Gamma^i_{ab}\, \partial_i \left(p^a p^b\right) &=& g^{ij} \left[ g_{tt} (\partial_i \dot{t}) (\partial_j \dot{t}) + 2g_{t\phi} (\partial_i \dot{t}) (\partial_j \dot{\phi})+ g_{\phi\phi} (\partial_i \dot{\phi}) (\partial_j \dot{\phi}) \right]\quad
\end{eqnarray}
Now, as shown in the four-dimensional spacetime \cite{LRStability}, $\partial_i V = \partial_i (p_a p^a)\, = \, p_a \partial_i  p^a \, = \, -E \,\partial_i \dot{t} + L \,\partial_i \dot{\phi} \, = \, 0$ and we can replace $\partial_i \dot{\phi} $ with $E/L\, \partial_i \dot{t}$. This simplifies the expression to
\begin{eqnarray}
   \Gamma^i_{ab}\, \partial_i \left(p^a p^b\right) &=&  g^{ij} (\partial_i \dot{t}) (\partial_j \dot{t}) \left[\, g_{\phi\phi} \sigma^2 +2 \sigma g_{t\phi} + g_{tt} \,\right]
\end{eqnarray}
where $\sigma=E/L$.
The term in the square bracket is proportional to the potential $V(x^i)$. In particular, it is equal to $-V(x^i)\,D/L^2$, where $D= g_{t\phi}^2-g_{tt} g_{\phi\phi}$. Since $V=0$ at LR, the term $\Gamma^i_{ab}\, \partial_i \left(p^a p^b\right)$ vanishes at LR. Therefore, at the location of LR, we get
\begin{eqnarray}
   \frac{1}{2}\partial_i \partial^i V &=& p^a p^b \,\partial_i  \Gamma^i_{ab} = R_{ab}\,p^a p^b
\end{eqnarray}
Then the Null energy condition at LR implies
\begin{eqnarray}
    T_{\mu\nu} p^\mu p^\nu &=&  Q\,  R_{ab}\,p^a p^b = Q\left( \,\frac{1}{2} \partial_i \partial^i V \,\right) \geq 0 \label{Stability Criteria in HD}
\end{eqnarray}
where $Q$ represents the theory-specific multiplying factors evaluated at the LR as given in Eqs. (\ref{Stability condition in fR}), (\ref{Stability condition in Scalar Tensor}) and (\ref{Stability condition in Horndeski}) in the modified theories of gravity mentioned in section \ref{stability in MOG}. The tensor $M^{ij}$ can then be identified as $(Q\, g^{ij})$ with NEC at LR given by $M^{ij} \mathcal{H}_{ij}\geq0$, where $\mathcal{H}_{ij}$ is the Hessian of potential $V$.

For simplicity, we first consider a diagonal metric on the $(d-2)$-dimensional subspace of non-Killing coordinates. For $Q>0$, the NEC at LR reduces to
\begin{eqnarray}
    \partial_i \partial^i V &=& \sum_i g^{ii} \partial_i^2 V \geq 0
\end{eqnarray}
As the $(d-2)$ non-Killing dimensions are spacelike, the metric components satisfy $g^{ii}>0$, which implies that when the NEC is satisfied, all components of $\partial_i^2 V$ cannot be negative at the LR. Hence, an extremum LR cannot be at a local maximum of the effective potential. Equivalently, the type \textbf{B} corotating LR and a type \textbf{A} counterrotating LR cannot exist.

We can now proceed to identify the allowed LR pairings directly in terms of the extrema or saddle points of the effective potential $V(x^i)$ in higher-dimensional ($d > 4$) spacetime. We find that in both odd- and even-dimensional spacetimes, corotating LRs emerge as pairs consisting of either (i) a local potential minimum and a saddle point with potential maximized along an odd number of directions, or (ii) two distinct saddle points with potential maximized along even and odd numbers of directions, respectively. For counterrotating LRs, the exact pairing depends on the spacetime dimensions. In even-dimensional spacetimes, LRs appear as pairs of either (i) a local potential minimum and a saddle point with potential minimized along an odd number of directions, or (ii) two saddle points with potential minimized along even and odd numbers of directions, respectively. Conversely, in odd-dimensional spacetimes, counterrotating LRs emerge as pairs of either (i) a local potential minimum and a saddle point with potential minimized along an even number of directions, or (ii) two saddle points with potential minimized along even and odd numbers of directions, respectively.

Therefore, unlike in four-dimensional general relativity, the existence of a stable LR is not guaranteed by the NEC in higher-dimensional spacetimes. In the UCO spacetime, where LRs appear as a pair of two types of saddle points of the potential, the absence of a stable LR eliminates the possibility of LR-associated mechanisms for nonlinear instability. This allows the consideration of horizonless UCOs as feasible black hole alternatives, although their overall dynamical stability requires a separate investigation.

When the metric on the $(d-2)$-dimensional non-Killing subspace is not diagonal, the argument can be generalized by exploiting the positive-definite nature of its metric. At an extremum of $V(x^i)$, the Hessian $H_{ij}=\partial_i\partial_j V$ is symmetric and transforms as a covariant rank-two tensor under non-singular local coordinate transformations in the $(d-2)$-dimensional subspace. Since the non-Killing directions are spacelike, one can always choose a local orthonormal frame at the LR in which the metric on this subspace takes the form $\bar{g}_{ij}=\delta_{ij}$. This construction is analogous to that of a local inertial frame, but is restricted to the $(d-2)$-dimensional spacelike subspace. In this frame, the NEC condition becomes
\begin{eqnarray}
    \bar{g}^{ij}\partial_i\partial_j V
    =\sum_i \partial_i^2 V \geq 0.
\end{eqnarray}
Consequently, the Hessian cannot be negative definite at the LR, and the extremum cannot be a local maximum of the potential, leading to the same conclusion as in the case of a diagonal metric.

In the stability analyses performed so far, the field equations explicitly include the Ricci tensor, which, upon contraction with the null vector $p^\mu$, resulted in a direct connection to the effective potential at the LR. The NEC at the LR constrains the Laplacian of $V$ with a proportionality factor $Q$, which is a scalar function. In theories wherein the Lagrangian contains higher-order curvature terms, such contraction may not always be expressible in terms of the effective potential or may lead to complex relations. In the following section, we show that, in the case of Einstein-Gauss-Bonnet theory, which is intrinsically higher dimensional, it is still possible to express the contraction through a straightforward relation with the potential. However, the NEC now contains the Hessian of the potential contracted with a curvature-dependent tensor.

\subsection{Stability in Einstein-Gauss-Bonnet (EGB) gravity}

EGB gravity is the first nontrivial case of the Lovelock theory \cite{Lanczos, Lovelock,LovelockReview}, with the unique extension of the Einstein-Hilbert action by terms quadratic in curvature that keeps the field equations second order in derivatives of the metric. It also arises as the leading curvature correction in the low-energy effective action of string theory \cite{Ghostfreestringtheories, StringGeneratedModel}. In four dimensions, the Gauss-Bonnet term is a total derivative and does not contribute to the field equations, and its effects therefore appear only for $d \geq 5$. The action for EGB theory in $d$ dimensions is given by
\begin{eqnarray}
    S &=& \frac{1}{2\kappa} \int d^d x \,\sqrt{-g}\; (R+\alpha_{GB}\,\mathcal{L}_{GB})
\end{eqnarray}
where $\alpha_{GB}$ is the Gauss-Bonnet coupling constant with dimensions of length squared and $\mathcal{L}_{GB}=R^2-4\, R_{\alpha\beta} R^{\alpha\beta}+ R_{\alpha\beta\mu\nu} R^{\alpha\beta\mu\nu}$ is the Gauss-Bonnet term. Varying the action with respect to the metric, we get the field equations of the theory as
\begin{eqnarray}
G_{\mu\nu}+\alpha_{GB} \,H_{\mu\nu} &=& \kappa T_{\mu\nu}
\end{eqnarray}
 where $G_{\mu\nu}$ is the Einstein tensor and the Gauss-Bonnet term leads to
\begin{equation}
H_{\mu\nu}=2\big(RR_{\mu\nu}-2R_{\mu\alpha}R^\alpha{}_\nu-2R_{\mu\alpha\nu\beta}R^{\alpha\beta}
+R_\mu{}^{\alpha\beta\gamma}R_{\nu\alpha\beta\gamma}\big)-\tfrac12 g_{\mu\nu}\mathcal{L}_{GB}
\end{equation}
Contracting with the null vector $p^\mu$ at LR gives
\begin{equation}
R_{\mu\nu} p^\mu p^\nu+\alpha_{GB}\, H_{\mu\nu} p^\mu p^\nu=\kappa\,T_{\mu\nu}p^\mu p^\nu
\end{equation}
The first term on the LHS is the GR term. We now proceed to express $H_{\mu\nu} p^\mu p^\nu$ at the LR in terms of the potential $V$. To simplify the proof, we use two null vectors $p^a$ and $\eta^a$ as basis vectors instead of the two Killing directions in the two-dimensional tangent space at the LR. The vector $p^a$ is chosen to be a tangent vector at the LR, while the vector $\eta^a$ is chosen such that $\eta^a \eta_a=0$ and $p^a \eta_a=-1$. For corotating (counterrotating) LR, $p^a=p^t(1,\Omega_p)$ with $\Omega_p=H_+(H_-)$ and $\eta^a=\eta^t\,(1,\Omega_n)$ with $\Omega_n=H_-(H_+)$ and $\eta^t=g_{\phi\phi}/2 D\,p^t$. The metric tensor can be written in terms of the basis $p^a$, $\eta^a$ as
\begin{equation}
g^{\mu\nu}=-p^\mu \eta^\nu-\eta^\mu p^\nu+g^{ij}\delta^\mu_i\delta^\nu_j
\end{equation}
and a vector component along basis $p^a$ and $\eta^a$ is written as $X_p=X_a p^a$ and $X_\eta=X_a \eta^a$. The new contravariant metric components $g^{pp}$, $g^{\eta\eta}$ vanish as $p^a$ and $\eta^a$ are null vectors and $g^{p\eta}=g^{\eta p}=-1$. The independent nonzero components of Riemann curvature in the new basis are $R_{p\eta p\eta}$, $R_{pipj}$, $R_{\eta i\eta j}$, $R_{pi\eta j}$ and $R_{ijkl}$. Components carrying an odd number of non-Killing indices vanish identically as a consequence of circularity and the two symmetries of spacetime. Now we can write the components of the Ricci tensor and the Ricci scalar in terms of the above Riemann tensor components at LR as
\begin{eqnarray}
    R_{pp} &=& g^{ij} R_{pipj}\label{Rpp}\\
    R_{p\eta } &=& R_{p\eta p \eta}+g^{ij} R_{pi\eta j}\label{Rpn}\\
    R_{\eta\eta} &=& g^{ij} R_{\eta i\eta j}\label{Rnn}\\
    R_{ij} &=& -R_{pi\eta j}- R_{\eta ipj}+ g^{kl} R_{kilj}\label{Rij}\\
    R &=&  -2 R_{p\eta p \eta} -4 g^{ij} R_{pi\eta j}+ g^{ij} g^{kl} R_{kilj}\label{R scalar}
\end{eqnarray}
 At LR the expression of $H_{\mu\nu}p^\mu p^\nu$ reduce to
 \begin{eqnarray}
    H_{ab}p^ap^b=H_{pp} &=& 2\,R \, R_{pp}-4\, R_{p\alpha}\, R^\alpha{}_p-4\, R_{p\alpha p\beta}\, R^{\alpha\beta}+2\, R_p{}^{\alpha\beta\gamma}\, R_{p\alpha\beta\gamma} \label{Hpp}
\end{eqnarray}
Using the definitions of Ricci tensor components and the Ricci scalar in Eqs. (\ref{Rpp}-\ref{R scalar}) and using the symmetries of Riemann tensor, we simplify the four terms in the above equation as
\begin{eqnarray}
    2\,R \, R_{pp}&=& (-4 R_{p\eta p \eta} -8 g^{ij} R_{pi\eta j}+ 2g^{ij} g^{kl} R_{kilj}) \, R_{pp}\label{EGB term1}\\
    -4\, R_{p\alpha}\, R^\alpha{}_p &=& 8 R_{pp} (R_{p\eta p \eta}+g^{ij} R_{pi\eta j}) \label{EGB term2}\\
    -4\, R_{p\alpha p\beta}\, R^{\alpha\beta}  &=& -4\,  R_{p\eta p\eta} \,R_{pp} -4 \, R_{pi pj}\, g^{ik} g^{jl} (-2R_{pk\eta l}+ g^{rs} R_{rksl}) \label{EGB term3}\\
    2\,  R_{p\alpha\beta\gamma}\,R_p{}^{\alpha\beta\gamma}  &=& -8 \, R_{pipj}\,g^{ik}g^{jl}\,R_{pk\eta l} \label{EGB term4}
\end{eqnarray}
Combining all the terms and separating the terms with $R_{pp}$ and $R_{pipj}$, we get
\begin{eqnarray}
     H_{pp} &=& 2\, R_{pp}\,(g^{ij} g^{kl} R_{kilj})-4 \, R_{pi pj}\, (g^{ik} g^{jl}\,g^{rs} R_{rksl})
\end{eqnarray}
We note here that the elimination of terms involving Riemann tensor components $R_{p\eta p\eta},\, R_{pi\eta j}$ and $R_{pk\eta l}$ is only possible due to the unique coefficients in the Gauss-Bonnet term. Any other combination of second-order curvature terms will not lead to such a compact form of $H_{pp}$, except a common multiplying number which can be absorbed into the Gauss-Bonnet coupling constant. The terms in the two brackets are the Ricci scalar and the Ricci tensor in the $(d-2)$-dimensional subspace of non-Killing coordinates evaluated at LR; we label them as $\mathcal{R}_{d-2}=g^{ij} g^{kl} R_{kilj}$ and $\mathcal{R}^{ij}_{d-2}=g^{ik} g^{jl}\,g^{rs} R_{rksl}$ respectively. Using this notation and the expression of $R_{pp}$ from Eq.(\ref{Rpp}), we can write
\begin{eqnarray}
    H_{pp} &=& -4\, (\mathcal{R}^{ij}_{d-2}-\frac{1}{2}\mathcal{R}_{d-2}\, g^{ij}) \, R_{pipj}= -4\, \mathcal{G}^{ij}_{d-2} \, R_{pipj}
\end{eqnarray}
where we have defined $\mathcal{G}^{ij}_{d-2}$ as the equivalent of the Einstein tensor in $(d-2)$-dimensional subspace of non-Killing coordinates. We note here that in previously considered theories, the stability criteria only included the trace $g^{ij}R_{pipj} = R_{pp}$, whereas in the EGB gravity, stability of LR depends on the individual components $R_{pipj}$. Now, using the expressions for nonzero Christoffel symbols given in the previous section, and considering the vanishing contribution from $g_{ab}(\partial_i p^a)(\partial_j p^b)=-V(x^i)\, D/L^2$ at LR in the expansion of $R_{pipj}=R_{aibj}p^a p^b$, we relate this component of the Riemann tensor to the effective potential as
\begin{eqnarray}
R_{pipj} &=& \frac{1}{2} \partial_j \partial_i V
\end{eqnarray}
Then the NEC at LR becomes
\begin{eqnarray}
    \kappa\,T_{pp} = R_{pp} +\alpha_{GB}\, H_{pp} &=& \left(g^{ij} -4\,\alpha_{GB}\, \mathcal{G}^{ij}_{d-2}\right) \,\left(\frac{1}{2} \partial_j \partial_i V\right)= \frac{1}{2}\mathcal{M}^{ij}\mathcal{H}_{ij}\geq 0 \quad \label{Stability criteria in EGB}
\end{eqnarray}
where $\mathcal{M}^{ij} = g^{ij} - 4\,\alpha_{GB}\, \mathcal{G}^{ij}_{d-2}$ and $\mathcal{H}_{ij} =\partial_i\partial_j V$ is the Hessian of the effective potential at the LR. For consistency, one can check that for $d = 4$, the tensor $\mathcal{G}^{ij}_{d-2}$ vanishes identically in the two-dimensional subspace and Eq.(\ref{Stability criteria in EGB}) reduces to the GR case as expected from the nature of the Gauss-Bonnet term in four dimensions. The first nontrivial case is $d=5$, where $G^{ij}_{3}$ is the Einstein tensor of the three-dimensional non-Killing subspace.

Although Eq.(\ref{Stability criteria in EGB}) does include an additional $\mathcal{G}^{ij}_{d-2}$ factor compared to the GR case, we can still comment on the stability of LRs using the tensor $\mathcal{M}^{ij}$ in the following three cases: (i) When $\mathcal{M}^{ij}$ is positive definite, the NEC is violated if the Hessian is negative definite. Hence, the LR cannot be a local maximum of the potential $V$. (ii) If $\mathcal{M}^{ij}$ is negative definite, a positive definite Hessian will lead to violation of the NEC, so the LR cannot be a local minimum of the potential $V$. (iii) In the case where $\mathcal{M}^{ij}$ is indefinite, the NEC constrains only a weighted trace of the Hessian, and no restriction on the type of the LR follows from Eq. (\ref{Stability criteria in EGB}) alone.
 
Note that the topological results of Section \ref{Pairs in HD} are independent of the field equations and therefore hold unchanged in EGB gravity. Only the restrictions on the stability of LRs are modified. We also emphasize here that EGB gravity differs qualitatively from the theories considered in previous sections. The Gauss-Bonnet correction to the NEC at the LR does not appear as an additional scalar factor with $g^{ij}$, as in Eq.(\ref{Stability Criteria in HD}), but as a curvature-dependent tensor contracted with the Hessian of the effective potential. The stability of the LR is then controlled by the interplay between the coupling constant and the local curvature at the LR, rather than by the sign of a single function. However, the conclusions of higher-dimensional GR persist whenever the GB coupling length scale is small compared with the curvature scale, say $l_c\sim |\mathcal{G}^{ij}_{d-2}|$, that is ($|\alpha_{GB}|\ll l_c$) at the LR.

The $m^{th}$ Lovelock gravity in $d$ dimensions contains the $m^{th}$ order higher-curvature terms with a special combination of the coefficients. This leads to a decomposition into surface and bulk terms, resulting in second-order field equations for $d> 2m$ \cite{Holography}. The general field equations are given as $E_{\mu\nu}=\kappa T_{\mu\nu}$ where $E_{\mu\nu}$ is the equation of motion (EOM) tensor \cite{EOMTensor}. The results for GR and EGB gravity suggest how the tensor $\mathcal{M}^{ij}$ may generalize to this family. In GR ($m=1$), the NEC at the LR contracts $R_{pipj}$ with $g^{ij}$, which is the zeroth-order EOM tensor of the non-Killing subspace. In EGB gravity, the $m=2$ term contracts $R_{pipj}$ with $\mathcal{G}^{ij}_{d-2}$, which is the first-order (Einstein) EOM tensor of the same subspace. It is therefore natural to expect that the $m^{th}$ order Lovelock term contributes $R_{pipj}$ contracted with the order-$(m-1)$ EOM tensor of the non-Killing subspace, $E_{(m-1),\,d-2}^{ij}$. If this holds, the NEC at the LR may retain the form $\mathcal{M}^{ij}\partial_i\partial_j V \geq 0$, with
\begin{equation}
  \mathcal{M}^{ij} = g^{ij} + \sum_{m\geq 2} \beta_k\,  E^{ij}_{(m-1),\,d-2}
  \label{M_Tensor_Lovelock}
\end{equation}
where the $\beta_k$ are numerical coefficients, with $\beta_2 = -4$ for EGB gravity. Such a tensor structure of $\mathcal{M}^{ij}$ will naturally reduce to the corresponding expressions in EGB and GR, thereby maintaining consistency with the known results. To establish the relation in Eq.(\ref{M_Tensor_Lovelock}), one needs to prove that the contributions of the remaining Riemann components, $R_{p\eta p\eta}$ and $R_{pi\eta j}$, cancel at every order, as they do for $m=2$. We expect the antisymmetry of the generalized Kronecker delta to enforce this elimination. We plan to develop the proof in future work.

In the case of a general quadratic gravity, the combination of higher-curvature terms may not lead to the second-order field equations, and their contraction with $p^\mu$ may lead to complex terms which may not reduce to an expression in terms of the effective potential. In such cases, one cannot associate NEC with LR stability. We demonstrate this by considering the following quadratic gravity action
\begin{eqnarray}
    S &=& \frac{1}{2\kappa}\int d^d x \sqrt{-g} \left(R+\psi_1\, R^2 + \psi_2 \,R_{\mu\nu} R^{\mu\nu}+\psi_3 \, R_{\alpha\beta\mu\nu} R^{\alpha\beta\mu\nu}\right)
\end{eqnarray}
where $\psi_1$, $\psi_2$ and $\psi_3$ are coupling constants. Varying the action with respect to the metric gives the following field equations \cite{EOM2}:
\begin{eqnarray}
     \kappa\, T_{\mu\nu} &=& G_{\mu\nu}-\frac{g_{\mu\nu}}{2} \left[\psi_1\, R^2+\psi_2\, R_{\alpha\beta} R^{\alpha\beta}+\psi_3 \, R_{\alpha\beta\gamma\delta} R^{\alpha\beta\gamma\delta}+(4\psi_1+\psi_2)\, \Box R\right]\nonumber\\
     & &+2\psi_1 R R_{\mu\nu}+ 2 \,(\psi_2+2\psi_3)\, R^{\alpha\beta} \, R_{\alpha\mu\beta\nu}-4\,\psi_3 \,R_{\mu\alpha}R^{\alpha}_{\,\nu}+2\,\psi_3 \,R_{\mu\alpha\beta\gamma} \, R_\nu^{\,\alpha\beta\gamma}\nonumber\\
     & &- (2\psi_1+\psi_2+2\psi_3)\, \nabla_\mu \nabla_\nu R +(\psi_2+4\psi_3)\,\Box R_{\mu\nu}
\end{eqnarray}
where $\Box=g^{\alpha\beta}\nabla_\alpha \nabla_\beta$ and $G_{\mu\nu}$ is the Einstein tensor. Now we find the contraction of the RHS of the above equation with null vectors $p^\mu p^\nu$. Using $g_{\mu\nu}p^\mu p^\nu=0$ for null vectors, $p^\mu p^\nu \nabla_\mu \nabla_\nu R=0$ at LR as shown in section \ref{$f(R)$ Gravity section} and Eqs. (\ref{EGB term1}-\ref{EGB term4}) we get
\begin{eqnarray}
    \kappa\, T_{\mu\nu} p^\mu p^\nu &=& R_{p p}+ 2\left[\psi_1 \,\mathcal{R}_{d-2} \,g^{ij}+(\psi_2+2\psi_3)\, \mathcal{R}^{ij}_{d-2} \right]R_{pi pj}\nonumber\\
    & &+(\psi_2+4\psi_3)\,p^\mu p^\nu\,\Box R_{\mu\nu}+2\,(-2\psi_1+ \psi_2+6\psi_3)\, R_{p\eta p\eta} \,R_{pp}\nonumber\\
    & &-4\,(\psi_2+4\psi_3)\, R_{pipj}\,g^{ik}g^{jl}\,R_{pk\eta l}+8(\psi_3-\psi_1) \, R_{pp} (g^{ij} R_{pi\eta j}) \label{null contraction quadratic}
\end{eqnarray}
One can simply check that for $(\psi_1,\psi_2,\psi_3)=(1,-4,1)$, the last four terms in the above equation vanish, and we recover the result for the EGB case. We emphasize again that this elimination is possible only due to the specific combination of Gauss-Bonnet coefficients and will not work for any other ratio of constants $(\psi_1,\psi_2,\psi_3)$. Now, expressing $R_{pp}$ and $R_{pipj}$ in terms of the effective potential, the NEC at LR implies
\begin{align}
    \left[1+2\,\psi_1 \,\mathcal{R}_{d-2} +2\,(-2\psi_1+ \psi_2+6\psi_3)\,R_{p\eta p\eta}+8(\psi_3-\psi_1) g^{ij} R_{pi\eta j} \right] g^{ij} \,&\partial_j \partial_i V\nonumber\\
    +\left[(\psi_2+2\psi_3)\mathcal{R}^{ij}_{d-2} -2(\psi_2+4\psi_3)\, g^{ik} g^{jl}R_{pk\eta l})\,\right]\,&\partial_j \partial_i V\nonumber\\
     \qquad \qquad \qquad + (\psi_2+4\psi_3)  \,p^\mu p^\nu\,&\Box R_{\mu\nu}\big|_{LR} \geq 0 \quad
\end{align}
Here, the first two terms can be expressed in terms of the second derivatives of the effective potential since they involve contraction of vectors $p^\mu$ with Riemann or Ricci tensor components. However, in the last term, the vectors $p^\mu$ are contracted with the covariant derivatives of the Ricci tensor, which cannot be related to the effective potential in a similar way. Therefore, one cannot straightforwardly comment on the stability of LR using NEC in such a quadratic theory of gravity and care must be taken to evaluate the extra terms at the LRs.

\section{Discussion} \label{Discussion}

The UCOs in four-dimensional GR have been argued to suffer from nonlinear instabilities associated with the accumulation of perturbations near stable LRs, raising questions about their viability as long-lived black-hole mimickers. It is therefore important to check whether the existence of stable LRs around smooth, horizonless UCOs is a universal feature of all gravitational theories in any number of dimensions or a property specific to four-dimensional GR. To address this question, we first investigated LR stability in modified theories of gravity in four dimensions, focusing on $f(R)$ gravity, scalar-tensor theories, and restricted Horndeski gravity. We then extended the analysis of the LR existence and stability theorem to higher-dimensional ($d>4$) stationary, axisymmetric spacetimes.

In four-dimensional modified gravity, we find that LR stability is governed not only by the effective potential of the spacetime but also by theory-dependent quantities. We identified a tensor $\mathcal{M}^{ij}$ that is contracted with the Hessian of the potential $\mathcal{H}_{ij}=\partial_i\partial_j V$ in the Null energy condition for each theory. This tensor provides the weights corresponding to each component of the Hessian $\mathcal{H}_{ij}$ and therefore provides a way to represent the stability criteria in any theory. The tensor $\mathcal{M}^{ij}$ was found to be $(df/dR) g^{ij}$, $(\Phi/\alpha) g^{ij}$, and $G_4(\Phi) g^{ij}$ for $f(R)$ theory, scalar-tensor theories, and restricted Horndeski's theory, respectively. In $f(R)$ gravity, the stability criterion depends on the sign of $df/dR$. For ghost-free $f(R)$ theories with $df/dR>0$, the stability criterion agrees with that in GR, and the UCO possesses a stable LR under the assumptions of the existence and stability analysis. For $f(R)=R$, our results consistently recover the GR case. Conversely, when $df/dR<0$, the extremum corresponds to a local maximum of the effective potential, and the corresponding LR is therefore locally unstable. Analogous conclusions follow in scalar-tensor theories and restricted Horndeski gravity, where the stability criteria depend on the respective factors $\Phi/\alpha$ and $G_4(\Phi)$. When these factors are positive, the LR is stable, whereas negative values correspond to an unstable LR. Thus, in the regimes where the relevant factors are negative, the stable LR need not exist, indicating that the LR-associated instability mechanism need not apply to such UCOs.

We next extended the LR existence and stability analysis to higher-dimensional stationary, axisymmetric spacetimes. By deriving the equations governing LRs and employing the Brouwer degree argument together with a stability analysis, we generalize the light ring existence theorem to higher dimensions. In $d$-dimensional spacetimes, an LR is characterized by a critical point specified by $(d-2)$ non-Killing coordinates, determined by a system of $(d-2)$ equations. We find that the topological argument underlying the pairwise occurrence of LRs persists in higher dimensions. However, the higher-dimensional vector-field map admits four distinct categories of LRs, including two types of saddle points. The two extremum categories correspond to local maxima and minima of the effective potential, while the two saddle categories, denoted saddle type I and saddle type II, correspond to the potential being maximum along an even and odd number of directions. Our stability analysis, using the NEC and the corresponding field equations of GR and modified theories of gravity, further constrains the possible LR pairs. The pairing varies for corotating and counterrotating LRs and also for odd or even numbers of dimensions. In particular, we find the LRs emerge in the following configurations: The corotating LRs appear as one of the following two pairs in both odd- and even-dimensional spacetime. The pairs are: (i) a locally stable LR and a saddle-point LR, with the potential being maximum along an odd number of directions, or (ii) two distinct saddle-point LRs, with the potential being maximum along even and odd numbers of directions, respectively. For counterrotating LRs, the pairing differs in even and odd-dimensional spacetime. In even-dimensional spacetimes, LRs appear as pairs of either (i) a locally stable LR and a saddle-point LR with potential being minimum along an odd number of directions, or (ii) two saddle-point LRs with potential being minimum along even and odd numbers of directions, respectively. While in odd-dimensional spacetimes, counterrotating LRs emerge as pairs of either (i) a locally stable LR and a saddle-point LR with potential being minimum along an even number of directions, or (ii) two saddle-point LRs with potential being minimum along even and odd numbers of directions, respectively.

Consequently, unlike in four-dimensional GR, the existence of a stable LR is not guaranteed by the NEC in higher dimensions, neither in GR nor in the modified theories considered. In particular, UCO configurations possessing LRs of two saddle types need not exhibit stable LRs. The absence of a stable LR eliminates the specific instability mechanism associated with the trapping of radiation or gravitational perturbations at a stable LR, although their overall nonlinear dynamical stability remains an open question.

We also analyzed LR stability in Einstein-Gauss-Bonnet gravity. In this case, the Gauss-Bonnet correction to the NEC at the LR takes the form $-4\alpha_{GB}\, \mathcal{G}^{ij}_{d-2} R_{pipj}$, where $\mathcal{G}^{ij}_{d-2}$ is the Einstein tensor of the non-Killing subspace. The NEC therefore constrains the Hessian of the effective potential weighted by the tensor $\mathcal{M}^{ij} = g^{ij} - 4\alpha_{GB}\, \mathcal{G}^{ij}_{d-2}$, rather than its Laplacian. This correction vanishes identically in four dimensions, consistent with the nature of the Gauss-Bonnet term. When $\mathcal{M}^{ij}$ is positive definite, which is guaranteed if the Gauss-Bonnet length scale is small compared with the curvature scale at the LR, the conclusions obtained for higher-dimensional GR remain unchanged. When $\mathcal{M}^{ij}$ is negative definite, the NEC instead forbids a locally stable LR. We have also discussed the expected form of the $\mathcal{M}^{ij}$ tensor in general $ m^ {th} $- order Lovelock theory, which we plan to develop in future work. We further examined the NEC in quadratic gravity and found that not all terms reduce to the form $\mathcal{M}^{ij}\mathcal{H}_{ij}\geq 0$, and hence it is not straightforward to comment on the stability of LR in the general quadratic theory of gravity.

Our results lead to three principal conclusions. First, the existence of a stable LR around a stationary, axisymmetric UCO is not a universal feature across gravitational theories and spacetime dimensions. Second, the absence of stable LRs in certain regimes of modified gravity and in higher-dimensional spacetimes opens the possibility of horizonless UCOs as long-lived black-hole mimickers. Whether such configurations can arise through gravitational collapse and remain dynamically viable requires further investigation. Third, since UCO solutions without stable LRs may also exist in higher-dimensional GR, the existence of a UCO, by itself, may not serve as a definitive test of GR.

It is instructive to note that our stability analysis does not require the spacetime to be horizonless, as it is local in nature and is performed in the neighborhood of a light ring. Consequently, the analysis also applies to light rings in the domain of outer communication of black hole spacetimes described by the metric ansatz in Eq. (\ref{metricSA}). In contrast, the presence of an event horizon can affect the existence analysis based on the topological argument, through the associated boundary conditions and global properties of the spacetime \cite{StationaryBHLRs}. Extending the existence analysis to higher-dimensional black hole spacetimes would be an interesting direction for future work.

\section*{Acknowledgment}
S. K.’s research is supported by the Anusandhan National Research Foundation, India, under the ANRF ARG Grant No. ANRF/ARG/2025/005821/PS.

\bibliographystyle{unsrt}
\bibliography{references}

@article{BosonStar,
doi = {10.1088/0264-9381/20/20/201},
url = {https://doi.org/10.1088/0264-9381/20/20/201},
year = {2003},
publisher = {},
volume = {20},
number = {20},
pages = {R301},
author = {Franz E Schunck and Eckehard W Mielke},
title = {General relativistic boson stars},
journal = {Classical and Quantum Gravity},
note={[arXiv:0801.0307]}
}

@article{Procastar,
title = {Proca stars: Gravitating Bose–Einstein condensates of massive spin 1 particles},
journal = {Physics Letters B},
volume = {752},
pages = {291-295},
year = {2016},
issn = {0370-2693},
doi = {https://doi.org/10.1016/j.physletb.2015.11.051},
url = {https://www.sciencedirect.com/science/article/pii/S0370269315009077},
author = {Richard Brito and Vitor Cardoso and Carlos A.R. Herdeiro and Eugen Radu},
note={[arXiv:1508.05395]}
}

@Article{Gravastar,
AUTHOR = {Mazur, Pawel O. and Mottola, Emil},
TITLE = {Gravitational Condensate Stars: An Alternative to Black Holes},
JOURNAL = {Universe},
VOLUME = {9},
YEAR = {2023},
NUMBER = {2},
ARTICLE-NUMBER = {88},
URL = {https://www.mdpi.com/2218-1997/9/2/88},
ISSN = {2218-1997},
DOI = {10.3390/universe9020088},
note={[arXiv:gr-qc/0109035]}
}

@book{Wormholes,
    author = "Visser, Matt",
    title = "{Lorentzian wormholes: From Einstein to Hawking}",
    isbn = "978-1-56396-653-8",
    year = "1995",
    publisher = "American Institute of Physics Melville, NY"
}

@article{LRStability,
  title = {Light-Ring Stability for Ultracompact Objects},
  author = {Cunha, Pedro V. P. and Berti, Emanuele and Herdeiro, Carlos A. R.},
  journal = {Phys. Rev. Lett.},
  volume = {119},
  issue = {25},
  pages = {251102},
  numpages = {6},
  year = {2017},
  publisher = {American Physical Society},
  doi = {10.1103/PhysRevLett.119.251102},
  url = {https://link.aps.org/doi/10.1103/PhysRevLett.119.251102},
  note={[arXiv:1708.04211]}
}

@article{StationaryBHLRs,
  title = {Stationary Black Holes and Light Rings},
  author = {Cunha, Pedro V. P. and Herdeiro, Carlos A. R.},
  journal = {Phys. Rev. Lett.},
  volume = {124},
  issue = {18},
  pages = {181101},
  numpages = {5},
  year = {2020},
  publisher = {American Physical Society},
  doi = {10.1103/PhysRevLett.124.181101},
  url = {https://link.aps.org/doi/10.1103/PhysRevLett.124.181101},
  note={[arXiv:2003.06445]}
}

@article{Universalproperties,
  title = {Universal properties of light rings for stationary axisymmetric spacetimes},
  author = {Guo, Minyong and Gao, Sijie},
  journal = {Phys. Rev. D},
  volume = {103},
  issue = {10},
  pages = {104031},
  numpages = {6},
  year = {2021},
  publisher = {American Physical Society},
  doi = {10.1103/PhysRevD.103.104031},
  url = {https://link.aps.org/doi/10.1103/PhysRevD.103.104031},
  note={[arXiv:2011.02211]}
}

@article{fRLivingReview,
  author = {De Felice, Antonio and Tsujikawa, Shinji},
  title = {$f({R})$ Theories},
  journal = {Living Reviews in Relativity},
  year = {2010},
  volume = {13},
  number = {1},
  pages = {3},
  doi = {10.12942/lrr-2010-3},
  url = {https://doi.org/10.12942/lrr-2010-3},
  note={[arXiv:1002.4928]}
}

@article{fRReviewModPhys,
  title = {$f({R})$ theories of gravity},
  author = {Sotiriou, Thomas P. and Faraoni, Valerio},
  journal = {Rev. Mod. Phys.},
  volume = {82},
  issue = {1},
  pages = {451--497},
  numpages = {0},
  year = {2010},
  publisher = {American Physical Society},
  doi = {10.1103/RevModPhys.82.451},
  url = {https://link.aps.org/doi/10.1103/RevModPhys.82.451},
  note={[arXiv:0805.1726]}
}

@article{scalartensortheories,
author = {Quiros, Israel},
title = {Selected topics in scalar–tensor theories and beyond},
journal = {International Journal of Modern Physics D},
volume = {28},
number = {07},
pages = {1930012},
year = {2019},
doi = {10.1142/S021827181930012X},
URL = {https://doi.org/10.1142/S021827181930012X},
eprint = {https://doi.org/10.1142/S021827181930012X},
note={[arXiv:1901.08690]}
}

@article{HorndeskiEOM,
    author = {Kobayashi, Tsutomu and Yamaguchi, Masahide and Yokoyama, Jun'ichi},
    title = {Generalized G-Inflation: Inflation with the Most General Second-Order Field Equations},
    journal = {Progress of Theoretical Physics},
    volume = {126},
    number = {3},
    pages = {511-529},
    year = {2011},
    note={[arXiv:1105.5723]},
    issn = {0033-068X},
    doi = {10.1143/PTP.126.511},
    url = {https://doi.org/10.1143/PTP.126.511},
    eprint = {https://academic.oup.com/ptp/article-pdf/126/3/511/6866310/126-3-511.pdf},
}

@article{HorndeskiGW170817,
author = {Kase, Ryotaro and Tsujikawa, Shinji},
title = {Dark energy in Horndeski theories after GW170817: A review},
journal = {International Journal of Modern Physics D},
volume = {28},
number = {05},
pages = {1942005},
year = {2019},
doi = {10.1142/S0218271819420057},
URL = {https://doi.org/10.1142/S0218271819420057},
eprint = {https://doi.org/10.1142/S0218271819420057},
note={[arXiv:1809.08735]}
}

@article{BardeenKerrorbits,
       author = {Bardeen, James M. and Press, William H. and Teukolsky, Saul A.},
        title = "{Rotating Black Holes: Locally Nonrotating Frames, Energy Extraction, and Scalar Synchrotron Radiation}",
      journal = {The Astrophysical Journal},
         year = {1972},
       volume = {178},
        pages = {347-370},
          doi = {10.1086/151796},
       adsurl = {https://ui.adsabs.harvard.edu/abs/1972ApJ...178..347B}
}

@article{Universalinterferometricsignatures,
author = {Michael D. Johnson et. al.},
title = {Universal interferometric signatures of a black hole’s photon ring},
journal = {Science Advances},
volume = {6},
number = {12},
pages = {eaaz1310},
year = {2020},
doi = {10.1126/sciadv.aaz1310},
URL = {https://www.science.org/doi/abs/10.1126/sciadv.aaz1310},
eprint = {https://www.science.org/doi/pdf/10.1126/sciadv.aaz1310},
note={[arXiv:1907.04329]}
}

@article{RingownprobeLR,
  title = {Is the Gravitational-Wave Ringdown a Probe of the Event Horizon?},
  author = {Cardoso, Vitor and Franzin, Edgardo and Pani, Paolo},
  journal = {Phys. Rev. Lett.},
  volume = {116},
  issue = {17},
  pages = {171101},
  numpages = {6},
  year = {2016},
  publisher = {American Physical Society},
  doi = {10.1103/PhysRevLett.116.171101},
  url = {https://link.aps.org/doi/10.1103/PhysRevLett.116.171101},
  note={[arXiv:1602.07309]}
}

@article{ShadowQNMcardoso,
  title = {Geodesic stability, Lyapunov exponents, and quasinormal modes},
  author = {Cardoso, Vitor and Miranda, Alex S. and Berti, Emanuele and Witek, Helvi and Zanchin, Vilson T.},
  journal = {Phys. Rev. D},
  volume = {79},
  issue = {6},
  pages = {064016},
  numpages = {13},
  year = {2009},
  publisher = {American Physical Society},
  doi = {10.1103/PhysRevD.79.064016},
  url = {https://link.aps.org/doi/10.1103/PhysRevD.79.064016},
  note={[arXiv:0812.1806]}
}

@article{RingdownandEchoes,
  title = {Gravitational-wave signatures of exotic compact objects and of quantum corrections at the horizon scale},
  author = {Cardoso, Vitor and Hopper, Seth and Macedo, Caio F. B. and Palenzuela, Carlos and Pani, Paolo},
  journal = {Phys. Rev. D},
  volume = {94},
  issue = {8},
  pages = {084031},
  numpages = {13},
  year = {2016},
  publisher = {American Physical Society},
  doi = {10.1103/PhysRevD.94.084031},
  url = {https://link.aps.org/doi/10.1103/PhysRevD.94.084031},
  note={[arXiv:1608.08637]}
}

@article{nonlinearinstability,
  title = {Light rings as observational evidence for event horizons: Long-lived modes, ergoregions and nonlinear instabilities of ultracompact objects},
  author = {Cardoso, Vitor and Crispino, Lu{\'{\i}}s C. B. and Macedo, Caio F. B. and Okawa, Hirotada and Pani, Paolo},
  journal = {Phys. Rev. D},
  volume = {90},
  issue = {4},
  pages = {044069},
  numpages = {10},
  year = {2014},
  publisher = {American Physical Society},
  doi = {10.1103/PhysRevD.90.044069},
  url = {https://link.aps.org/doi/10.1103/PhysRevD.90.044069},
  note={[arXiv:1406.5510]}
}

@article{LRsin5D,
  title = {Light rings around five dimensional stationary black holes and naked singularities},
  author = {Tavlayan, Aydin and Tekin, Bayram},
  journal = {Phys. Rev. D},
  volume = {107},
  issue = {2},
  pages = {024016},
  numpages = {12},
  year = {2023},
  publisher = {American Physical Society},
  doi = {10.1103/PhysRevD.107.024016},
  url = {https://link.aps.org/doi/10.1103/PhysRevD.107.024016},
  note={[arXiv:2209.14873]}
}

@article{Kajol,
  title = {Photon region boundary for stationary axisymmetric black holes},
  author = {Padhye, Prasad and Paithankar, Kajol and Kolekar, Sanved},
  journal = {Phys. Rev. D},
  volume = {112},
  issue = {12},
  pages = {124025},
  numpages = {17},
  year = {2025},
  publisher = {American Physical Society},
  doi = {10.1103/b84x-ch1m},
  url = {https://link.aps.org/doi/10.1103/b84x-ch1m},
  note={[arXiv:2412.12665]}
}

@article{ngEHT,
  author = {Ayzenberg, Dimitry et. al.},
  title = {Fundamental physics opportunities with the next-generation Event Horizon Telescope},
  journal = {Living Reviews in Relativity},
  year = {2025},
  volume = {28},
  number = {1},
  pages = {4},
  doi = {10.1007/s41114-025-00057-0},
  url = {https://doi.org/10.1007/s41114-025-00057-0},
  note={[arXiv:2312.02130]}
}

@inproceedings{BHEXphoton,
author = {Alexandru Lupsasca et. al.},
title = {{The Black Hole Explorer: photon ring science, detection, and shape measurement}},
volume = {13092},
booktitle = {Space Telescopes and Instrumentation 2024: Optical, Infrared, and Millimeter Wave},
organization = {International Society for Optics and Photonics},
publisher = {SPIE},
pages = {130926Q},
year = {2024},
doi = {10.1117/12.3019437},
URL = {https://doi.org/10.1117/12.3019437},
note={[arXiv:2406.09498]}
}

@article{TeoSPOs,
  author = {Teo, Edward},
  title = {Spherical Photon Orbits Around a Kerr Black Hole},
  journal = {General Relativity and Gravitation},
  year = {2003},
  volume = {35},
  number = {11},
  pages = {1909--1926},
  doi = {10.1023/A:1026286607562},
  url = {https://doi.org/10.1023/A:1026286607562},
  note={[arXiv:2007.04022]}
}

@article{Horndeski1974,
  author = {Horndeski, Gregory Walter},
  title = {Second-order scalar-tensor field equations in a four-dimensional space},
  journal = {International Journal of Theoretical Physics},
  year = {1974},
  volume = {10},
  number = {6},
  pages = {363--384},
  doi = {10.1007/BF01807638},
  url = {https://doi.org/10.1007/BF01807638}
}

@article{GW170817Paper1,
  title = {Strong Constraints on Cosmological Gravity from GW170817 and GRB 170817A},
  author = {Baker, T. and Bellini, E. and Ferreira, P. G. and Lagos, M. and Noller, J. and Sawicki, I.},
  journal = {Phys. Rev. Lett.},
  volume = {119},
  issue = {25},
  pages = {251301},
  numpages = {6},
  year = {2017},
  publisher = {American Physical Society},
  doi = {10.1103/PhysRevLett.119.251301},
  url = {https://link.aps.org/doi/10.1103/PhysRevLett.119.251301},
  note={[arXiv:1710.06394]}
}

@article{GW170817Paper2,
  title = {Dark Energy after GW170817 and GRB170817A},
  author = {Creminelli, Paolo and Vernizzi, Filippo},
  journal = {Phys. Rev. Lett.},
  volume = {119},
  issue = {25},
  pages = {251302},
  numpages = {6},
  year = {2017},
  publisher = {American Physical Society},
  doi = {10.1103/PhysRevLett.119.251302},
  url = {https://link.aps.org/doi/10.1103/PhysRevLett.119.251302},
  note={[arXiv:1710.05877]}
}

@article{GW170817Paper3,
  title = {Dark Energy After GW170817: Dead Ends and the Road Ahead},
  author = {Ezquiaga, Jose Mar{\'{\i}}a and Zumalac{\'a}rregui, Miguel},
  journal = {Phys. Rev. Lett.},
  volume = {119},
  issue = {25},
  pages = {251304},
  numpages = {6},
  year = {2017},
  publisher = {American Physical Society},
  doi = {10.1103/PhysRevLett.119.251304},
  url = {https://link.aps.org/doi/10.1103/PhysRevLett.119.251304},
  note={[arXiv:1710.05901]}
}

@article{Timedelay,
  title = {Time delay as a probe of multiple photon spheres},
  author = {Paithankar, Kajol and Kolekar, Sanved},
  journal = {Phys. Rev. D},
  volume = {114},
  issue = {8},
  pages = {084001},
  numpages = {19},
  year = {2026},
  publisher = {American Physical Society},
  doi = {10.1103/7qv6-xrc7},
  url = {https://link.aps.org/doi/10.1103/7qv6-xrc7},
  note={[arXiv:2604.14132]}
}

@article{Wong2024,
doi = {10.3847/2041-8213/ad8650},
url = {https://doi.org/10.3847/2041-8213/ad8650},
year = {2024},
publisher = {The American Astronomical Society},
volume = {975},
number = {2},
pages = {L40},
author = {Wong, George N. and Medeiros, Lia and Cárdenas-Avendaño, Alejandro and Stone, James M.},
title = {Measuring Black Hole Light Echoes with Very Long Baseline Interferometry},
journal = {The Astrophysical Journal Letters},
note={[arXiv:2410.10950]}
}

@article{LUAshadowcorrespondence,
  title = {Black hole shadow and acceleration bounds for spherically symmetric spacetimes},
  author = {Paithankar, Kajol and Kolekar, Sanved},
  journal = {Phys. Rev. D},
  volume = {108},
  issue = {10},
  pages = {104042},
  numpages = {14},
  year = {2023},
  publisher = {American Physical Society},
  doi = {10.1103/PhysRevD.108.104042},
  url = {https://link.aps.org/doi/10.1103/PhysRevD.108.104042},
  note={[arXiv:2305.07444]}
}

@article{HOD2018,
title = {On the number of light rings in curved spacetimes of ultra-compact objects},
journal = {Physics Letters B},
volume = {776},
pages = {1-4},
year = {2018},
issn = {0370-2693},
doi = {https://doi.org/10.1016/j.physletb.2017.11.021},
url = {https://www.sciencedirect.com/science/article/pii/S0370269317309139},
author = {Shahar Hod},
note={[arXiv:1710.00836]}
}

@article{XavierHerdeiroCrispino2024,
  title = {Traversable wormholes and light rings},
  author = {Xavier, S{\'e}rgio V. M. C. B. and Herdeiro, Carlos A. R. and Crispino, Lu{\'{\i}}s C. B.},
  journal = {Phys. Rev. D},
  volume = {109},
  issue = {12},
  pages = {124065},
  numpages = {10},
  year = {2024},
  publisher = {American Physical Society},
  doi = {10.1103/PhysRevD.109.124065},
  url = {https://link.aps.org/doi/10.1103/PhysRevD.109.124065},
  note={[arXiv:2404.02208]}
}

@article{GhoshSarkar2021,
  title = {Light rings of stationary spacetimes},
  author = {Ghosh, Rajes and Sarkar, Sudipta},
  journal = {Phys. Rev. D},
  volume = {104},
  issue = {4},
  pages = {044019},
  numpages = {5},
  year = {2021},
  publisher = {American Physical Society},
  doi = {10.1103/PhysRevD.104.044019},
  url = {https://link.aps.org/doi/10.1103/PhysRevD.104.044019},
  note={[arXiv:2107.07370]}
}

@article{Wei2020,
  title = {Topological charge and black hole photon spheres},
  author = {Wei, Shao-Wen},
  journal = {Phys. Rev. D},
  volume = {102},
  issue = {6},
  pages = {064039},
  numpages = {12},
  year = {2020},
  publisher = {American Physical Society},
  doi = {10.1103/PhysRevD.102.064039},
  url = {https://link.aps.org/doi/10.1103/PhysRevD.102.064039},
  note={[arXiv:2006.02112]}
}

@article{Babichev2014Dressing,
  author = {Babichev, Eugeny and Charmousis, Christos},
  title = {Dressing a black hole with a time-dependent Galileon},
  journal = {Journal of High Energy Physics},
  year = {2014},
  volume = {2014},
  number = {8},
  pages = {106},
  doi = {10.1007/JHEP08(2014)106},
  url = {https://doi.org/10.1007/JHEP08(2014)106},
  note={[arXiv:1312.3204]}
}

@article{M87,
doi = {10.3847/2041-8213/ab0ec7},
url = {https://doi.org/10.3847/2041-8213/ab0ec7},
year = {2019},
publisher = {The American Astronomical Society},
volume = {875},
number = {1},
pages = {L1},
author = {The Event Horizon Telescope Collaboration},
title = {First M87 Event Horizon Telescope Results. I. The Shadow of the Supermassive Black Hole},
journal = {The Astrophysical Journal Letters},
note={[arXiv:1906.11238]}
}

@article{SgrA,
doi = {10.3847/2041-8213/ac6674},
url = {https://doi.org/10.3847/2041-8213/ac6674},
year = {2022},
title={First Sagittarius {A}* Event Horizon Telescope Results. I. The Shadow of the Supermassive Black Hole in the Center of the Milky Way},
publisher = {The American Astronomical Society},
volume = {930},
number = {2},
pages = {L12},
author = {The Event Horizon Telescope Collaboration},
journal = {The Astrophysical Journal Letters},
note={[arXiv:2311.08680]}
}

@inproceedings{BHEXmotivation,
author = {Michael D. Johnson et. al.},
title = {{The Black Hole Explorer: motivation and vision}},
volume = {13092},
booktitle = {Space Telescopes and Instrumentation 2024: Optical, Infrared, and Millimeter Wave},
editor = {Laura E. Coyle and Shuji Matsuura and Marshall D. Perrin},
organization = {International Society for Optics and Photonics},
publisher = {SPIE},
pages = {130922D},
year = {2024},
doi = {10.1117/12.3019835},
URL = {https://doi.org/10.1117/12.3019835},
note={[arXiv:2406.12917]}
}

@article{LIGOfirst,
  title = {Observation of Gravitational Waves from a Binary Black Hole Merger},
  author = {LIGO Scientific Collaboration and Virgo Collaboration},
  journal = {Phys. Rev. Lett.},
  volume = {116},
  issue = {6},
  pages = {061102},
  numpages = {16},
  year = {2016},
  publisher = {American Physical Society},
  doi = {10.1103/PhysRevLett.116.061102},
  url = {https://link.aps.org/doi/10.1103/PhysRevLett.116.061102},
  note={[arXiv:1602.03837]}
}

@article{blindtotopology,
  title = {Light rings on stationary axisymmetric spacetimes: Blind to the topology and able to coexist},
  author = {Cunha, Pedro V. P. and Herdeiro, Carlos A. R. and Novo, Jo{\~a}o P. A.},
  journal = {Phys. Rev. D},
  volume = {109},
  issue = {6},
  pages = {064050},
  numpages = {17},
  year = {2024},
  publisher = {American Physical Society},
  doi = {10.1103/PhysRevD.109.064050},
  url = {https://link.aps.org/doi/10.1103/PhysRevD.109.064050},
  note={[arXiv:2401.05495]},
}

@article{MOGReview1,
title = {Modified gravity theories on a nutshell: Inflation, bounce and late-time evolution},
journal = {Physics Reports},
volume = {692},
pages = {1-104},
year = {2017},
issn = {0370-1573},
doi = {https://doi.org/10.1016/j.physrep.2017.06.001},
url = {https://www.sciencedirect.com/science/article/pii/S0370157317301527},
author = {S. Nojiri and S.D. Odintsov and V.K. Oikonomou},
note={[arXiv:1705.11098]}
}

@article{MOGReview2,
title = {Unified cosmic history in modified gravity: From F(R) theory to Lorentz non-invariant models},
journal = {Physics Reports},
volume = {505},
number = {2},
pages = {59-144},
year = {2011},
issn = {0370-1573},
doi = {https://doi.org/10.1016/j.physrep.2011.04.001},
url = {https://www.sciencedirect.com/science/article/pii/S0370157311001335},
author = {Shin’ichi Nojiri and Sergei D. Odintsov},
note={[arXiv:1011.0544]}
}

@article{Lanczos,
 ISSN = {0003486X, 19398980},
 URL = {http://www.jstor.org/stable/1968467},
 author = {Cornelius Lanczos},
 journal = {Annals of Mathematics},
 number = {4},
 pages = {842--850},
 publisher = {[Annals of Mathematics, Trustees of Princeton University on Behalf of the Annals of Mathematics, Mathematics Department, Princeton University]},
 title = {A Remarkable Property of the Riemann-Christoffel Tensor in Four Dimensions},
 urldate = {2026-11-10},
 volume = {39},
 year = {1938}
}

@article{Lovelock,
    author = {Lovelock, David},
    title = {The Einstein Tensor and Its Generalizations},
    journal = {Journal of Mathematical Physics},
    volume = {12},
    number = {3},
    pages = {498-501},
    year = {1971},
    issn = {0022-2488},
    doi = {10.1063/1.1665613},
    url = {https://doi.org/10.1063/1.1665613},
    eprint = {https://pubs.aip.org/aip/jmp/article-pdf/12/3/498/19155957/498_1_online.pdf},
}

@article{Ghostfreestringtheories,
title = {Curvature squared terms and string theories},
journal = {Physics Letters B},
volume = {156},
number = {5},
pages = {315-317},
year = {1985},
issn = {0370-2693},
doi = {https://doi.org/10.1016/0370-2693(85)91616-8},
url = {https://www.sciencedirect.com/science/article/pii/0370269385916168},
author = {Barton Zwiebach}
}

@article{StringGeneratedModel,
  title = {String-Generated Gravity Models},
  author = {Boulware, David G. and Deser, S.},
  journal = {Phys. Rev. Lett.},
  volume = {55},
  issue = {24},
  pages = {2656--2660},
  numpages = {0},
  year = {1985},
  month = {Dec},
  publisher = {American Physical Society},
  doi = {10.1103/PhysRevLett.55.2656},
  url = {https://link.aps.org/doi/10.1103/PhysRevLett.55.2656}
}

@article{Holography,
  title = {Holography in action},
  author = {Kolekar, Sanved and Padmanabhan, T.},
  journal = {Phys. Rev. D},
  volume = {82},
  issue = {2},
  pages = {024036},
  numpages = {12},
  year = {2010},
  month = {Jul},
  publisher = {American Physical Society},
  doi = {10.1103/PhysRevD.82.024036},
  url = {https://link.aps.org/doi/10.1103/PhysRevD.82.024036},
  note={[arXiv:1005.0619]}
}

@article{EOMTensor,
  author = {Kolekar, Sanved},
  title = {On the Bianchi identity in generalized theories of gravity},
  journal = {General Relativity and Gravitation},
  year = {2022},
  volume = {54},
  number = {8},
  pages = {92},
  doi = {10.1007/s10714-022-02978-5},
  url = {https://doi.org/10.1007/s10714-022-02978-5},
  isbn = {1572-9532}
}

@article{LovelockReview,
title = {Lanczos–Lovelock models of gravity},
journal = {Physics Reports},
volume = {531},
number = {3},
pages = {115-171},
year = {2013},
issn = {0370-1573},
doi = {https://doi.org/10.1016/j.physrep.2013.05.007},
url = {https://www.sciencedirect.com/science/article/pii/S0370157313002615},
author = {T. Padmanabhan and D. Kothawala},
note = {[arXiv:1302.2151]}
}

@article{EOM2,
  title = {Cosmological and black hole brane-world universes in higher derivative gravity},
  author = {Nojiri, Shin'ichi and Odintsov, Sergei D. and Ogushi, Sachiko},
  journal = {Phys. Rev. D},
  volume = {65},
  issue = {2},
  pages = {023521},
  numpages = {22},
  year = {2001},
  publisher = {American Physical Society},
  doi = {10.1103/PhysRevD.65.023521},
  url = {https://link.aps.org/doi/10.1103/PhysRevD.65.023521},
  note={[arXiv:hep-th/0108172]}
}

\end{document}